\documentclass[twocolumn,
reprint,
amsmath,amssymb,
floatfix
]{revtex4-2}
\usepackage{natbib}
\usepackage{graphicx}
\usepackage{dcolumn}
\usepackage{bm}
\usepackage{float}
\usepackage{aas_macros}
\usepackage{xcolor}

\newcommand\numberthis{\addtocounter{equation}{1}\tag{\theequation}}
\newcommand{\Salamanca}{Department of Fundamental Physics and IUFFyM, University of Salamanca,\\ Plaza de la Merced S/N E-37008, Salamanca, Spain}
\begin{document}

\title{Gravitational Wave emission in Binary Neutron Star \\early post-merger within a dark environment}

\author{D. Suárez-Fontanella}
\email{duvier@usal.es}
\author{D. Barba-González}
\email{david.barbag@usal.es}
\author{C. Albertus}
\email{albertus@usal.es}
\author{M. \'Angeles P\'erez-Garc\'ia}
\email{mperezga@usal.es}

 \affiliation{\Salamanca}

\date{\today}

\begin{abstract}
Using an effective Lagrangian model inspired by Takami et al. \cite{takamiPhysRevD.91.064001} we qualitatively study  the early post-merger of a nearly symmetric binary Neutron Star (BNS) merger event with a non-vanishing ambient fraction of dark matter. For this we first mimic the dynamics of two oscillating Neutron Star (NS) masses in the gravitational potential well as they merge. We parametrize the dynamics and ejecta properties in the coalescence event allowing the formation of a surrounding debris disk that may be containing a non-vanishing dark matter fraction.
In order to analyze the possible novel dark contribution, we start from a dark-matter free modellization as a benchmark.  Using Monte Carlo Markov Chain (MCMC) techniques we approximately recover the  gravitational waveforms, restricted to  early post-merger time interval from existing simulations in the CoRe database. Later, we explore the impact of an additional  dark viscous fluid under a prescribed velocity dependent force in the Lagrangian and obtain the resulting waveforms and some spectral features originating in the first few ms in the BNS post-merger. Finally we discuss our qualitative findings and its range of validity in light of the prospects of detectability in present or future experimental settings.
\end{abstract}

\keywords{dark matter,  effective lagrangian, gravitational waves, binary neutron star merger}

\maketitle
\section{Introduction}
\label{sec:Introduction}
The detection of dark matter (DM) is one of the most exciting and enigmatic current challenges in astrophysics, cosmology and particle physics. On-going searches of this elusive type of matter include different strategies. Namely they are mostly associated to scattering events (direct searches), production of secondaries such as photons or neutrinos (indirect searches) or production in hadron colliders and beam-dump experiments at high energies (collider searches), see \cite{Arbey:2021gdg} for a review. There is now a vast literature on the phenomenology associated to the, otherwise, feebly interacting dark sector \cite{Marsh:2024ury}.  Current models predict that based on prescribed interactions with ordinary matter, up to $\sim10\%$ mass fraction can reside as  DM in the Neutron Star (NS) stellar volume \cite{Mariani_2023} and beyond. Halo distributions where radial extent of DM is beyond the baryonic value, $R_\chi>R_{b}$, are also possible for light and weakly bound DM candidates \cite{2024Parti...7..201R}.

So far different particle candidates have been proposed to account for the DM. Ranging from weakly to strongly interacting species and for a mass range nowadays spanning many orders of magnitude, there are plenty of models in the literature. Just to cite a few of them, let us remind here the weakly interacting massive particle (WIMP) such as the supersymmetric neutralino, asymmetric dark matter (ADM), light bosons such as axions or axion-like particles and strangelets  \cite{Oks_2021}. Although much theoretical and experimental efforts have been undertaken, its nature remains elusive, constituting an open problem. Nevertheless, there is now overwhelming evidence of its existence more than nine decades after the pioneering observations by Fritz Zwicky in 1933 \cite{zwicky1933rotverschiebung}, who showed that the velocities of galaxies in the Coma cluster greatly exceeded the expectations based solely on the sum of the individual luminous galaxy masses. Since then, wide searches involving different strategies trying to hunt  these exotic particles have shown null results although the phase space is now more  tightly constrained.

Remarkably, the dawn of the era of multimessenger physics allows to gather information on different channels enriching the picture and more efficiently constraining the physical mechanisms governing its behaviour. In particular,  gravitational waves (GWs) experiments are expected to provide a precise reconstruction of the properties of inspiraling and merging compact objects, including mass/radius/composition measurements as they would be sensitive to tiny deviations in the gravitational waveforms which may be induced by nobel  features in the violent events of merging black holes (BHs) and NSs. These environmental effects could be the smoking-gun to signal the presence of DM, although this requires a careful identification of the ordinary baryon content \cite{Bertone_2020}.

Since the LIGO first detection in 2015 of GWs emitted during the merger of a stellar-mass binary black hole (BBH) event \cite{ligobbh}, the possibility to add this new multimessenger opens a new window to the Universe. Further, GWs provide a new channel to shed light on the nature of DM, also constituting a definite confirmation that BBHs exist and merge with a local rate \cite{rate} in the range between 17.9 and $44 \hspace{2pt}\mathrm{Gpc}^{-3} \mathrm{yr}^{-1}$ at a fiducial redshift $(z=0.2)$, assuming the known sources are representative of the total population. Other compact binary coalescence (CBC) search targets include collisions of BHs and NSs or binary NS mergers (BNS), where in all of them GWs are produced in a transient emission that can be detected on current terrestrial facilities. Recently, detectors including LIGO-Virgo-Kagra (LVK) have released enough data to build catalogs currently listing nearly 100 detections \cite{Nitz_2023}. Specifically the  Gravitational-Wave Transient Catalog 3 (GWTC-3) \cite{rate} contains signals consistent with three classes of binary mergers. They infer the BNS merger rate to be between 10 and $1700 \hspace{2pt}\mathrm{Gpc}^{-3} \mathrm{yr}^{-1}$ and the NS-BH merger rate to be between 7.8 and $140\hspace{2pt} \mathrm{Gpc}^{-3} \mathrm{yr}^{-1}$, assuming a constant rate density in the comoving frame and taking the union of $90 \%$ credible intervals. Besides these efforts, other future 3rd-generation planned detectors such as Cosmic Explorer (CE) \cite{evans2023cosmicexplorersubmissionnsf}, Einstein Telescope \cite{Branchesi_2023}, NEMO \cite{nemoAckley_2020} or LISA \cite{amaroseoane2017laserinterferometerspaceantenna}  will be able to further detect ultralow and ultraweak GW spectral sensitivities of the order $h\sim 10^{-25}$ $\rm Hz^{-1/2}$ produced by massive cosmic events with redshifts $z\lesssim 10$ and total mass around that expected in BNS  $M \lesssim3 M_\odot$. 


In this work we will use a Lagrangian effective model inspired by Takami et al \cite{takamiPhysRevD.91.064001, ELLIS2018607} in order to qualitatively assess possible effects in the dynamics of the early post-merger phase in a BNS merger occurring in an environment polluted by viscous DM. Thus in our model the BNS environment is characterized by a baryonic mass fraction $f_b$ and DM fraction $f_\chi$ during the GW emission time interval. We compute the set of polarization strengths $h_{+,\times}(t)$ of the GW signal as a function of time $t$ and subsequent observables derived from them. 
We note that, at present, there are refined tools such as Bilby \cite{Ashton_2019}, a Bayesian inference library for GW astronomy, allowing to infer the source astrophysical properties from existing experimental data. This tool does not incorporate, however,  particular realizations of exotic models including DM like the one we present in this work. 

In our setting the presence of DM in the system is due to the actual BNS merger having place in a DM polluted  environment. It may be due to stripped DM from the inside of the NS or from an overdensity or halo where typically DM density can be much larger with respect to the solar neighborhood value $\rho_\chi\simeq0.4$ $\rm GeV/cm^3$. It has been recently claimed \cite{garani2023JCAP...07..038G} that even overdensities of a few  $\sim 10^4$ $\rm GeV/cm^3$ at 1 parsec from the center of M4 would not yield definite constraints on DM from kinetically heating of NSs or white dwarfs.

The detection of DM using GW along with different strategies has already been discussed, see \cite{Bertone_2020}. One of the key ideas is based on the fact that it constitutes a fingerprint of  the formation and evolution of first cosmic structures. Popular models include cold DM  (CDM) on sub-galactic scales. However, there still remain incompatibilities between the CDM description and the observed data, as it predicts more structure on small scales than what we actually observe. Other models such as warm or fuzzy DM, as shown by simulations, lead to the suppression of small-scale structures and this, in turn, may lead to a significant reduction in the rate of BBH mergers observed, helping to further constrain among existing DM models. On the small scales, collisionless DM fluid faces major issues such as the core cusp problem, missing satellites problem or too big to fail problem, see \cite{2015PNAS..11212249W}. This shows that future GW observations will provide a new probe of physics beyond the $\Lambda$CDM model \cite{PhysRevD.108.043512}.

The possible traces that DM candidates could leave on the patterns of GW polarizations has been a particularly effervescent topic in the last decade \cite{fornalPhysRevD.108.055022,Samanta_2022,mosbePhysRevD.108.055022,Banerjee_2023, ghoshal2023probing}. The existence of viscous and self-interacting DM (SIDM) is another key point.  At the small scale, the SIDM behaves like a collisional dark matter but on the large scale it behaves like the collisionless DM, thus fitting expectations. In \cite{Mishra2022-ii} it is pointed out that the collisional nature of DM can lead to viscosity and using the kinetic theory, viscous coefficients of the SIDM can be obtained. Therefore the ratio of the SIDM cross section to its mass $(\sigma/m_\chi)$, for the present observed cosmic acceleration, can be reconciled with the constraints obtained from astrophysical observations.  Note that a GW will also experience the damping effect when it propagates in a fluid with non-zero viscosity so that the constraints of the luminosity distances from the observed GW events by experiments can be directly used to set constraints on self-interactions \cite{lu2018damping,gwviscodoi:10.1142/S0218271819501335}.


The structure of this contribution is as follows. In section \ref{sec:model} the Lagrangian model inspired by \cite{takamiPhysRevD.91.064001} is introduced along with the relevant set of effective parameters used to approximately reproduce the set of GW polarizations selected in our study from the  Computational Relativity (CoRe) database \cite{Dietrich_2018} involving several equations of state (EoS).
In section \ref{sec:modelA}, the different polarization modes are fit, first to a DM-free model, in order to assess the efficiency of the Monte Carlo Markov Chain (MCMC) procedure allowing for a parameter estimation and, later on, allowing for a non-vanishing viscous DM fraction. We discuss the obtained parameter values and provide a physical insight of them.

In section \ref{sec:PSD} we calculate the power spectral density (PSD) for the full signal obtained from the selected dataset, set as a benchmark, in order to  compare to that restricting to our post-merger results. This is done in order to inspect and qualitatively characterize the possible GW features and the sensitivity of current and future detectors to novel dark components, based on the relative effective signal-to-noise ratio for different DM polluted environments. Finally in section \ref{sec:Conclusions} we conclude.

\begin{figure}[th]
	\centering	\includegraphics[width=1.0\linewidth]{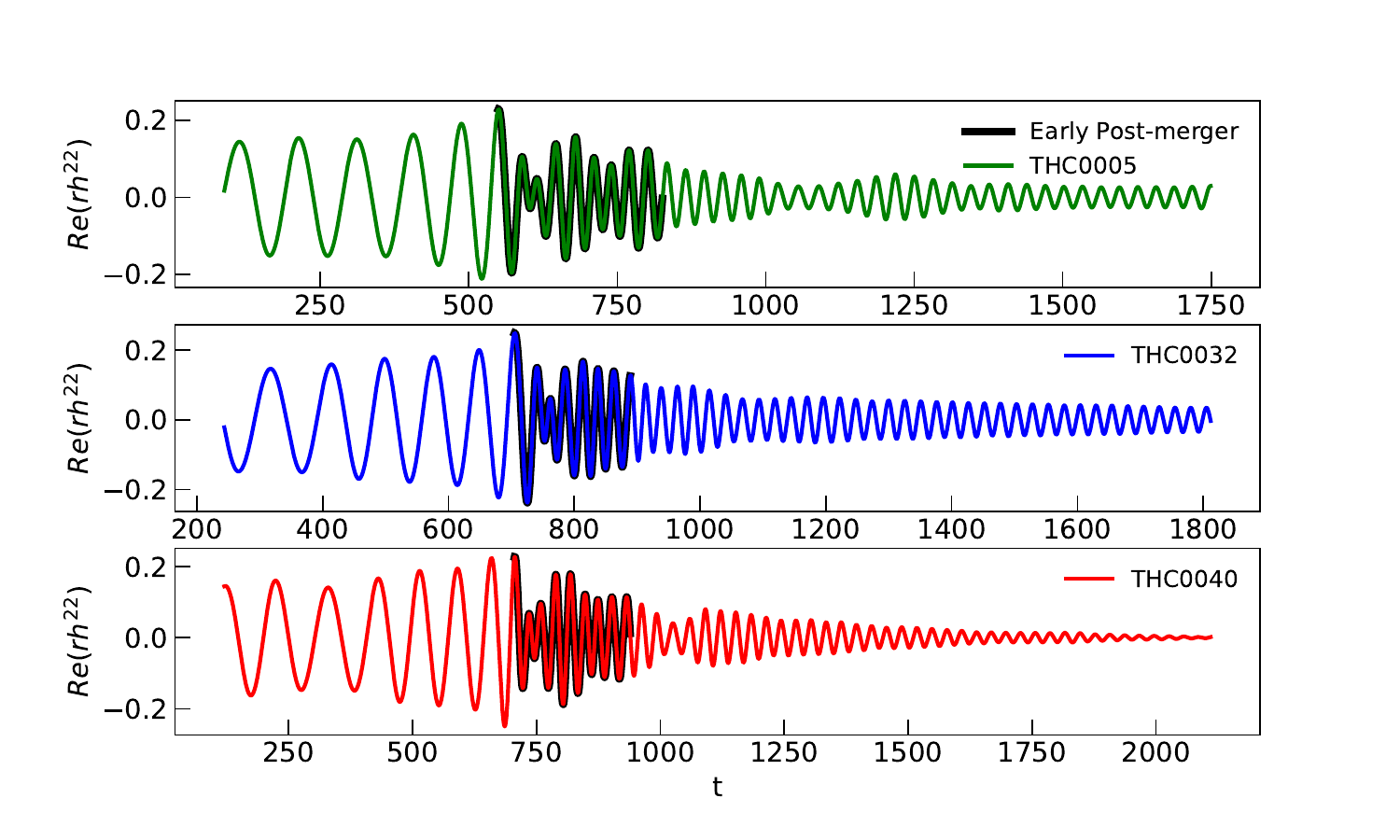}
	\caption{Principal mode $h^{22}$ data as a function of time from three simulations (THC0005, THC0032, THC0040) from CoRe database \cite{Dietrich_2018} obtained using WhiskyTHC. Our fitting dataset is limited to the early post-merger stage (highlighted in dark in each waveform). Time axis is expressed in $G M_{\odot} / c^3 \simeq 4.92549 \times 10^{-6} \mathrm{~s}$ units. }
	\label{fig:datas}
\end{figure}

\section{Model for the emission of Gravitational waves}
\label{sec:model}

The system we will scrutinize are the first few ms right after the NSs merge. Note that in this phase, strong tidal forces have already disrupted the otherwise nearly spherical (rotating) stellar configuration in asymptotically early times before the merger. 
In our scenario we will assume that a dark polluted environment is present due to a non-vanishing fraction of DM coexisting with the expected baryonic matter mass fraction expelled in the collision. This dark fraction origin may well come either from a liberated dark population loosely bound in superficial stellar layers or beyond baryonic radii, such as in halo configurations, see \cite{Mariani_2023} or from an existing local DM overdensity if the BNS merger takes place inside a DM cluster. 

As computational simulations show \cite{annurev:/content/journals/10.1146/annurev-nucl-013120-114541,bnssimPhysRevD.110.024003, Nedora_2021, Depietri2020PhRvD.101f4052D, Topolski_2024} just to cite a few,  for the few ms duration the BNS early post-merger dynamics show the presence of tidal forces and in particular oscillations of the compact objects in the common gravitational potential well, characterized by the lapse function. These simulations involve complex  microphysics under strong and electroweak scales in a curved space-time with dimensionality effects.

 
Since a technical description seems a very complex task we focus on a more affordable goal of qualitatively exploring a scenario where we study how a dark fluid composed of massive $\chi$ particles may leave a fingerprint in the transient early post-merger emission of a BNS merger event. For this we will use a minimally coupled DM model. In this respect DM should be thermalized with the NS stellar matter so that it can (at least partially) follow the dynamics of the merger. This condition is fulfilled for DM models yielding constant scattering spin-independent $\chi$-nucleon cross-sections   for $m_\chi\gtrsim 1$ GeV$/\rm c^2$, $\sigma_{\chi,N}\gtrsim 10^{-45}$ $\rm cm^2$, see \cite{Bell_2024}. Further couplings at the strong (hadronic) force limit are theoretically possible yielding a more ultradense environment \cite{herreroPhysRevD.100.103019} driving the final object to a BH collapse. However, we will restrict the DM sector we consider to yet another fraction of matter component, essentially inert with respect to changes in ordinary baryonic matter species.

\subsection{Binary Neutron Star post-merger dynamical oscillations}
\label{sec:modelA}

As mentioned, we will approximately describe the early post-merger dynamical behavior as that of an spinning oscillator. A word of caution is in order regarding the scenario depicted. According to previous works \cite{KiuchiPhysRevD.101.084006} this model can reasonably describe BNS nearly symmetric where one of the main peaks $f_1$ in the spectrogram of post-merger gravitational waves is interpreted as the spin frequency.  However, as claimed, this peak cannot be clearly identified for the asymmetric binary systems because the less massive NS is tidally disrupted before the merger and there are no prominent double dense cores. We assume in our calculation that the two cores survive as such during the post-merger phase i.e. the asymmetry is not large.

In that setting we will generally consider the effects of dissipation among two fluids (baryon/dark) and that from self-interacting DM. We will use an effective viscous (dissipative) force  $F_{d}\sim v_r^n$ under the form of a power of radial velocity $v_r=\dot{r}$, see below. Calculations in other scenarios such as the global absorption signal of 21 cm line at $z\sim17$ in the cosmic dawn era as reported in \cite{Bhatt_2019} propose a term to account for this gas-DM scattering terms affecting energy exchange among the two fluids $\sim\frac{d Q_C}{d t}$. Further, in order to have sufficient cooling of the gas at the cosmic dawn era, the DM-gas interaction is quoted to be sufficiently large, within existing constraints. We will not enter into these refinements in our model but we will consider it is non-vanishing. 
On general grounds in case the DM self-interaction takes the form $\sigma(v)=\sigma_0 v^{-\alpha}$ the typical classical force in the fluid can be written as $
F\sim n v p \sigma(v)=n p \sigma_0 v^{1-\alpha}$ where $n\equiv n_\chi, p$ are density and momentum transferred in each scattering. In our setting we will restrict to integer $\alpha=[-1,1]$ but in works where BBH mergers are simulated \cite{GonzaloPhysRevLett.133.021401} this parameter is explored, being key to facilitate the drag and collapse.

Specifically, in the case of the environment following a BNS merger, the viscosity of nuclear matter seems to be  relevant at early times $\lesssim3$ms after merger when timescales of local density oscillations and weak interactions are similar. Different neutrino microscopic approaches at the typical resolutions considered for mergers yield differences not largely affecting the early post-merger waveform, see \cite{Zappa_2023}.The system may undergo local density oscillations while similarly undergoing local changes in the fluid chemical composition \cite{espino2023neutrinotrappingoutofequilibriumeffects}.

In our scenario we will consider that the merger dynamics are affected by this non-ideal behavior with an effective viscosity from (dark/baryon) fluids including a self-interaction of the DM fraction in the form $F_{d} \simeq C_v \dot{r}^n$ with coefficients  $C_v\equiv \alpha_n, \beta_n\, (n=1,2)$ to be discussed later. 

Using MCMC we will fit the oscillation mode data as a function of time, obtained in simulations such as those in CoRe database \cite{Dietrich_2018} for a set of three different EoS, namely BHBlp \cite{2014ApJS..214...22B}, SFHo \cite{steiner2013core}, and DD2 \cite{2010PhRvC..81a5803T}. 
We note the dark fluid we incorporate  within our Lagrangian description will nevertheless constitute a small mass fraction of the total merger mass. As discussed, the ultimate source of DM viscosity is an assimilated non-zero self-interaction and non-zero  $\chi$-nucleon  cross-section that although weak in strength may be, however, similar to photon-photon scattering in electromagnetic plasmas for energies or hundredths of MeV.

More specifically matter will be expelled from the merging NS outer layers, less gravitationally bound, due to the tidal forces. In our setting we will consider this fraction of ejected matter forms a disk around the NS nuclei providing the scenario for Kilonova emission, being detectable with Earth based telescopes such as MAAT \cite{prada2020whitepapermaatgtc} in IR-UV bands in the neutron-rich environment formed after the merger. 

In this scenario we consider two merging NSs with gravitational masses $m_1$ and $m_2$ in a rotating disk with a fraction of DM, $f_\chi$, and an ordinary i.e. baryonic fraction, $f_b$,  whose origin is the debris of the tidal disruption of the two compact stars. In the mechanical analog from \cite{takamiPhysRevD.91.064001}, both spheres are connected by a spring characterized by constant $k$. The spheres can oscillate along the radial coordinate and rotate on the two-dimensional orbiting plane. For the sake of simplicity, we will consider an observer perpendicular to the plane where the orbital motion happens. As the damping oscillation proceeds, there will be a dynamical variation of the moment of inertia of the system and, from the conservation of angular momentum, this will be resulting in the generation of GW with a specific waveforms.

Note that specifically in  \cite{takamiPhysRevD.91.064001,ELLIS2018607} they are describing, however,  a different case i.e. a four-mass oscillator where two DM and two NS matter cores oscillate jointly with two different springs with constants $k_1,k_2\neq k_1$. Our model is different as it considers the fact that the oscillations happen in a dark environment and aims at qualitatively exploring features observed in the early post-merger signal. As we will describe,  these oscillations will be affected by the presence of DM and this, in turn, determines any further changes.

Despite its simplicity, this effective model can reasonably mimic the early BNS merger signal. It provides a first-step to improve BNS early post-merger signal description, potentially enabling exploring features of GW signals in environments including a DM component.\\

In order to evaluate the sensitivity of our model to DM in the predicted GW signal strength we will keep restricted to a minimum the number of extra parameters describing the dark environment where the BNS merger happens. 
Note that this effectively non-relativistic parametrization is based on the fact that there may be non-vanishing cross-sections among the two sectors and a possibly self-interacting term in the dark sector. As we will show, the non-relativistic nature of DM dynamics can be included in our minimal scenario in a meaningful way.

{As mentioned above, the Lagrangian we consider is able to mimic the GW polarizations of the BNS early post-merger signal. 
Thus the BNS merger event can be thought of in this scenario as two phases. In a first phase of the merger the outer NS layers are stripped off due to the tidal fields, carrying away the baryonic mass $m_b=f_{\text{ejec}}(m_1+m_2)\equiv f_{\rm ejec}M_T$ from the progenitor stars ($m_1, m_2$) rapidly forming a disk-like structure with a radius $R_d$, enclosing the two NS inner nuclei of masses $(1-f_{\text{ejec}})m_1$ and $(1-f_{\text{ejec}})m_2$, respectively bound by an elastic force of constant $k$. We assume the mass ratio is such that the two nuclei actually survive as such during the early post-merger time window of a few ms. We use equivalently $f_{\text{ejec}}\equiv f_b$. The minimum elongation happens at a characteristic radial distance, $l$, from the coordinate origin. 
Due to the abrupt fall of the tidal field after the ejection of the NS outer layers and the conservation of angular momentum, these nuclei keep rotating  inside this pre-collapsed object. 
Later on, presumably a complete fusion follows inside (second phase) as soon as the rotational energy of these nuclei decays that, however, we will not study.

For now, focusing on the interest of this work, we will restrict ourselves to the first of these phases, during which a fraction $f_\chi=m_\chi/M_T$ of DM  is present in the disk from the environment. Both  $f_b,f_\chi$ are small fractions at the per-cent level, as numerical relativity simulations seem to indicate \cite{bezaresPhysRevD.100.044049,bnssimPhysRevD.110.024003}.
We set the origin of coordinates at the center of mass of the merging NSs and use as dynamical  variables the distance from the center of coordinates to one of the spheres and its angle $(r, \theta)$ to describe the two degrees of freedom of the system.

For a system of mass ratio $q=m_1/m_2$ $(q\gtrsim 1)$, we can write the following Lagrangian as an {\it ideal} starting point 

\begin{align}\label{Lag}
	\mathcal{L}&=(1-f_{\rm ejec})q\frac{M_T}{2}(\dot{r}^{2}+ (r\dot{\theta})^{2}) + \nonumber \\  &(f_{\rm ejec}+f_{\chi})\frac{M_T}{2}R_d^{2}\dot{\theta}^{2}- \frac{k}{2}(1+q^{2})\left(r-\frac{l}{1+q}\right)^{2}.
\end{align}}

Note that the chirp mass may also be used as it is straightforwardly related to the total mass as  
\begin{equation}
\mathcal{M}_c=\left[\frac{q}{(1+q)^2}\right]^{3 / 5} M_T,
\end{equation}
and quite precisely measured in BNS mergers. In our modellization we have assumed that the angular velocity in the disk $\Omega\simeq \dot{\theta}$. We would like to remark  here that, as simulations show, see for example \cite{KastaunPhysRevD.91.064027},  differential rotation appears inside the disk. There and after a dozen of ms outer layers at $\sim 15$ km are rotating slightly below Keplerian velocity, while the core  rotates more slowly within factors of order unity. However we expect the addition of a viscous effect due to a net fraction of DM will drive even this approximation more realistic and, more importantly, not invalidate our results. 

{
We now proceed to generalize the previous idealized scenario to include DM-induced  dissipation. Depending on the conditions of the neutrino transport in BNS mergers dense matter may exhibit relevant microscopic viscous effects. Viscosities  for numerous EoS have been thoroughly studied in the literature \cite{Glampedakis_2018,Dall_Osso_2018}. We will discuss on this point later on when data sets we use are presented. Models for DM may also include a  viscous character by construction, as discussed in Section \ref{sec:Introduction} and recently shown in BBH simulations where viscous DM facilitates supermassive black holes enough to drop them within a parsec of each other \cite{GonzaloPhysRevLett.133.021401}. 

For the sake of concreteness we will consider the dissipative force $F_d$ in the Lagrangian dynamics can be generically described as a radial force proportional to $n$-power of radial velocity $v_r^n=\dot{r}^n$ with a density-dependent coefficient, proportional to the linear density.

Due to the different nature of both sectors and combining the baryonic and DM contributions we will set the effective dissipative force as
\begin{align}\label{disforce}
	F_d&=-\alpha_2 \frac{m_b}{R_d}\dot{r}^{2}-\beta_1 \frac{m_\chi}{R_d}\dot{r}-\beta_2 \frac{m_\chi}{R_d}\dot{r}^2,
\end{align}
where $\alpha_2,\beta_1, \beta_2$ are the generic dissipative coefficients. $\beta_{1,2}$ are induced by DM interaction while we choose to set a non-zero  $\alpha_2$ term for baryon. Effects from neutrino leakage scheme and viscous ordinary matter seem to not largely distort the GW waveforms \cite{Zappa_2023} and in that sense we have verified this from our results, see below.  In an alternative expression we obtain

\begin{equation}
    F_d=-[f_{\rm ejec}\alpha_2 \dot{r}^{2} +f_\chi(\beta_1 \dot{r}+\beta_2 \dot{r}^{2})] \frac{M_T}{R_d}.
\end{equation}

We can now apply the Euler-Lagrange formalism to obtain the equations of motion 

\begin{align}\numberthis\label{moveq}
	&\begin{aligned}&\ddot{r}+\frac{(\alpha_2 f_{\rm ejec}+\beta_2 f_{\chi})}{(1-f_{\rm ejec})}\frac{2}{qR_d}\dot{r} -	  \dot{\theta}^{2}r+\\&\nonumber\frac{(1+q^{2})}{(1-f_{\rm ejec})}\frac{k}{\xi}(r-\frac{l}{1+q})+ \beta_1\frac{f_\chi}{(1-f_{\rm ejec})}\frac{1}{qR_d} =0,	\end{aligned}\\\nonumber
 &(r^{2}+\frac{f_{\rm ejec}+f_{\chi}}{1-f_{\rm ejec}}\frac{R_d^{2}}{q})\ddot{\theta}+2 r\dot{r}\dot{\theta}=0,
\end{align} 
where $\xi\equiv qM_T$.

{\subsection{GW Polarizations}
To compute the GW polarizations, $h_{\times},h_{+}$ we employ the time variation of the reduced quadrupole moment $Q_{ij}=I_{ij}-\frac{1}{3}\delta_{ij}\delta^{\mu\nu}I_{\mu\nu} $ with $i,j=1,2,3$ in Cartesian coordinates. $I_{ij}, \delta_{ij}$ are the moment or inertia and Kronecker delta, respectively. With these definitions, we can write \cite{takamiPhysRevD.91.064001,thorne2000gravitation}:
\begin{align}\label{pol1}
	h_{+}&=\frac{\ddot{Q}_{11}-\ddot{Q}_{22}}{d},\hspace{5pt}
	h_{\times}=\frac{2\ddot{Q}_{12}}{d}
\end{align}
where $ d $ represents the distance to an optimally oriented i.e face-on observer.  
Using that
\begin{equation}\label{I}
	I_{ij}=\frac{M_T}{2}r^{2}\tilde{I}_{ij},
\end{equation}
where $\tilde{I}_{ij}$ is
\begin{equation}
	\begin{pmatrix}
		1+\rm cos(2\theta) & \rm sin(2\theta)\\
		\rm sin(2\theta) & 1-\rm cos(2\theta)\\
	\end{pmatrix},
\end{equation} 
and since $Q_{11}=\frac{M_T}{2}r^{2}(\frac{1}{3}+\rm cos(2\theta))$, $Q_{22}=\frac{M_T}{2}r^{2}(\frac{1}{3}-\rm cos(2\theta))$ and $Q_{12}=\frac{M_T}{2}r^{2}\rm sin(2\theta)$ it finally results

\begin{align}\label{gwpolarizaciones1}
h_+\, d&=A(r,\theta)\mathrm {cos}(2\theta)-B(r,\theta) \mathrm {sin}(2\theta),\\
h_{\times}\,d&=B(r,\theta)\mathrm{cos}(2\theta)+A(r,\theta)\rm sin(2\theta)\label{gwpolarizaciones2}.
\end{align}
Functions $A(r,\theta)$ and $B(r,\theta)$ can be obtained from the Lagrangian in Eq. (\ref{Lag}) as
\begin{align}
    A(r,\theta)&=2(1-f_{\rm ejec})qM_T(\dot{r}^{2}+r\ddot{r}-2r^{2}\dot{\theta}^{2})\\
    B(r,\theta)&=2(1-f_{\rm ejec})qM_T(4r\dot{r}\dot{\theta}+r^{2}\ddot{\theta})
\end{align}

Note that the ambient DM present during the merger is characterized by a non-vanishing fraction $f_\chi \neq0$, although we will consider cases where it has either viscous or inviscid nature. This is the minimal assumption that we have introduced as we aim to probe its potential implications on the polarization shape of the waves and spectral properties. More in-depth DM characterization would involve a non-trivial mixing involving force carriers among the two sectors that we will leave for a future work.

To elucidate these effects, we will proceed as follows. First, we will calibrate the DM-free model by obtaining a full model parameter set reasonably adjusting waveforms in the early BNS post-merger from selected  numerical relativity simulations. Later on, we will allow a restricted non-zero set of $f_\chi, \beta_{1,2}$ in order to determine the impact on the results. We will be considering three pivotal scenarios. One where DM exhibits no self-interaction viscosity ($f_\chi\neq 0, \beta_1=\beta_2=0$), a second one where it showcases effective viscosity with baryons, albeit lower by a factor $\sim 1/10$  than the previous, sourced from stars ($f_\chi\neq 0, \beta_{1,2}< \alpha_2$), and lastly, a scenario where self-interaction viscosity is similar or larger than DM-baryon ($f_\chi\neq 0, \beta_2 \gtrsim \alpha_2$).}  To accomplish these tasks, we will employ fitting methodologies grounded in Monte Carlo methods implemented in Python language, see below.

{\subsection{Fitting data sets}\label{datasection}
For our study, we use the CoRe database  \cite{Dietrich_2018} of BNS simulations restricted to a few ms in the early post-merger. The region of interest to our calculation is depicted in Fig.(\ref{fig:datas}) highlighted with a dark line for each waveform. Time axis is expressed in $G M_{\odot} / c^3 \simeq 4.92549 \times 10^{-6} \mathrm{~s}$ units. From top to bottom we plot the principal mode data $h^{22}$ from three simulations in the catalog (THC0005, THC0032, THC0040) obtained using WhiskyTHC \cite{2012A&A...547A..26R}.

Specifically, we extracted simulation data from THC0005, THC0032, and THC0040, employing the set of BHBlp \cite{2014ApJS..214...22B}, SFHo \cite{steiner2013core}, and DD2 \cite{2010PhRvC..81a5803T} EoS, respectively. The selected dataset shows some spread regarding initial BNS configurations, grids and use different neutrino schemes: neutrino leakage (LK) for the former two and a more advanced LK+M0 (hereafter M0) for the latter. A key point is neutrino trapping so that, if it occurs, modified-Urca processes can lead to strong bulk viscous dissipation and to damping \cite{Alford_2018, Alford_2022} of the remnant density oscillations possibly leaving a signature in the post-merger GW signal. More advanced schemes, such as M1, found  no significant out-of-thermodynamic equilibrium effects on the post-merger. \cite{Radice_2022}. Thus M1 simulations do not show evidences of enhanced damping of the radial oscillations of the remnant compared to the M0 runs. This suggests that the impact of bulk viscosity cannot be too large. It is however necessary to perform multi-resolution studies in more depth to elucidate this.

Dataset THC0005 Fig.(\ref{fig:datas}) top panel, relates to a BNS simulation with progenitor masses $m_1 = 1.4 M_\odot$ and $m_2 = 1.2 M_\odot$, $q=1.166$. Stellar structure equations associate for the BHBlp EoS their approximate radii as $R_{1,2} \approx 13.0$ km \cite{2014ApJS..214...22B}.

THC0032 in Fig.(\ref{fig:datas}) middle panel accurately reproduces the GW signature generated by two NSs with masses $m_1=1.4M_\odot$, $m_2=1.2M_\odot$, corresponding to approximate radii of $R_1= R_2= 11.50$ km. Similarly, for THC0040 Fig.(\ref{fig:datas}) lower panel, NS masses are $m_1=1.432M_\odot$, $m_2=1.3M_\odot$, $q=1.1055$ with radii $R_1=R_2=13.0$ km. In our setup we fix somewhat arbitrarily an approximate disk radius $R_d \simeq 15.44$ km.\\
From simple inspection, we see additional information regarding this GW mode, summarized in the shapes depicted in Fig.(\ref{fig:datas}). We will analyze it further in the framework of the full model when the DM component is incorporated, see below. The quoted values of mass and radius from study cases in simulations will serve as initial sampling points within the parameter space of our model.
\begin{figure}[t]
	\centering
	\includegraphics[width=1.0\linewidth]{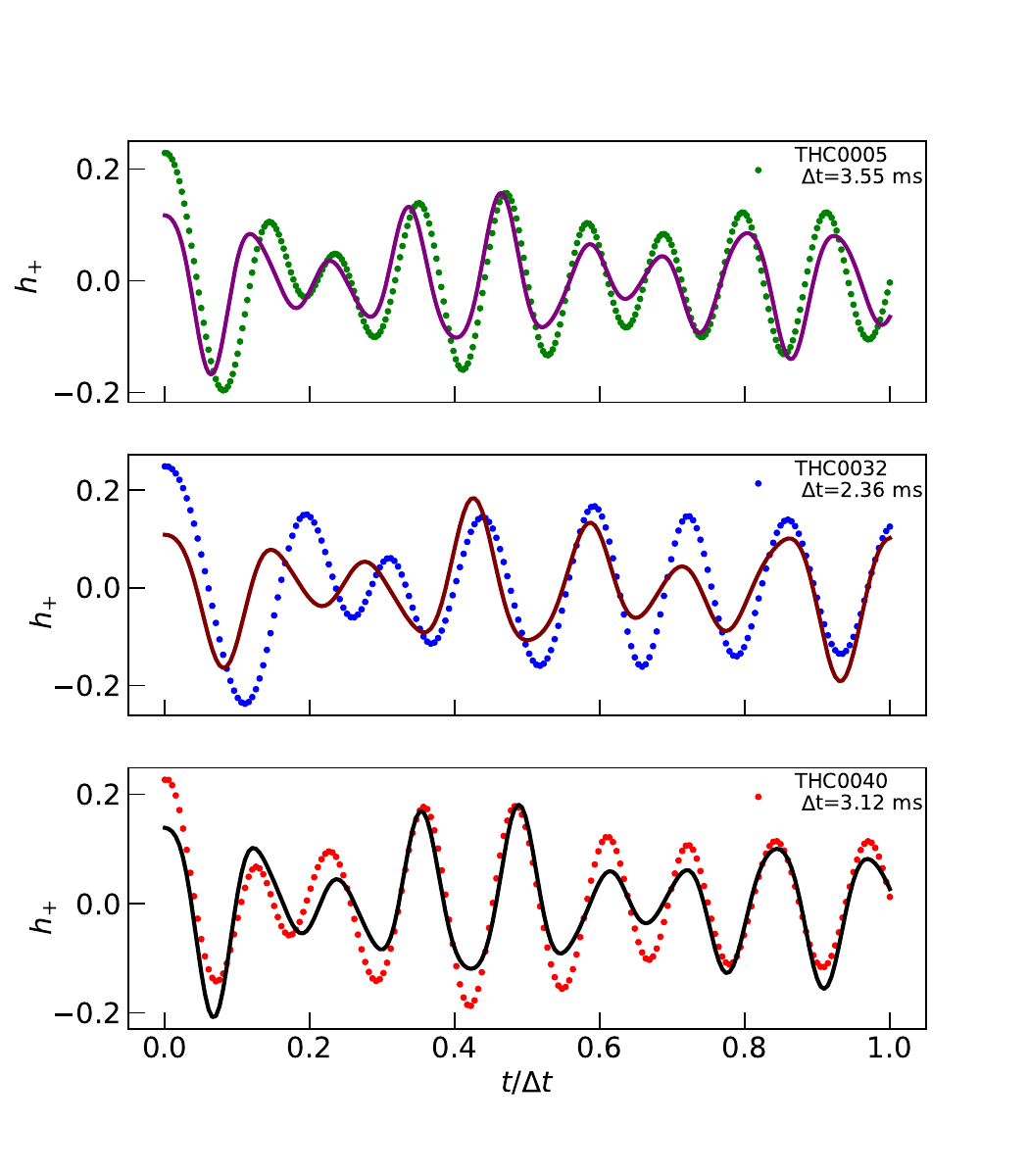}
	\caption{BNS gravitational waveform  from CoRe data base (points) and GW polarizations $h_+$ (solid lines) as computed from Eq.(\ref{pol1}). Fitting parameters appear in Table (\ref{T:vectores}). From top to bottom panel we depict the post-merger time interval $\Delta_t=3.55$ ms for THC0005 (green points), $\Delta_t=2.36$ ms for THC0032 (blue points) and $\Delta_t=3.12$ ms for THC0040 (red points). Time axis is scaled as  $t/\Delta t$ for each case. $h_{+}$ is rescaled by a $10^{-22}$ factor.}
	\label{fig:Sol}
\end{figure}

In the model presented, the parameter space encompasses the 10-tuple $(k, \alpha_2, \omega_0,  r_0, l, \tilde{f}, m_1, m_2, f_\chi, \beta_n)$ selecting $n=1,2$ or $\beta_n\equiv 0$ and $\omega(t_{\rm merger})\equiv\omega_0$, $r(0)\equiv r_0$. Each parameter holds a distinct physical significance. Specifically, $k$ is the effective spring constant of the elastic force defined for the gravitational potential in the post-merger. $\alpha_2, \beta_n$ are associated with dissipative forces in Eq. (\ref{disforce}) for the ordinary (dark) sector. $\omega_0$ is the initial angular velocity of the system, intricately linked to that of the binary system at the moment of merger. $r_0$ stands for the initial separation distance between the cores, while the parameter $l$ represents the length where the elastic force between the cores reaches its minimum value. The parameter $\tilde{f}$ stems from   $f_{\text{ejec}} \equiv \tilde{f} \frac{M_T}{\overline{R}}$ and ${\overline{R}}\equiv \frac{1}{2}\sum_{i=1,2} R_i$. This redefinition of the ejected baryonic mass  parameter partly accounts for the compactness of the mass-radius relation of the merging NSs and the associated EoS (supposed to be the same for both NSs). 


\subsection{Setup}\label{MCMC setup}
Using the CoRe database selected cases and the parameter space described in the section \ref{datasection}, we proceed to fit the proposed DO model. To this end, we employ the open-source emcee module in Python \cite{Foreman_Mackey_2013} which implements the Metropolis-Hastings method as a variant of the Markov Chain Monte Carlo algorithm   \cite{Foreman_Mackey_2013}.
Using MCMC we explore the parameter space spanned by the 10-tuple of physical quantities associated with the model, and minimize the distance between CoRe simulation provided main mode $h^{22}$ and the output of the equations of motion in our model shown in Eq.(\ref{moveq}) from which the solutions $r(t),\theta(t)$ obtained with a Runge-Kutta method are used to construct the GW polarization modes $h_+(t)$ and $h_\times(t) $ in the form Eq.(\ref{gwpolarizaciones1})  and Eq.(\ref{gwpolarizaciones2}). 

The simulation data consist of the early post-merger stage of the GW normalized in the temporal coordinate. This normalization significantly facilitates the search for parameter space; however, it must be taken into account when attempting to recover the values of the frequencies associated with this stage. So setting $t_{\rm merger}=0$ the time intervals for each of the data sets THC0005, THC0032, and THC0040 will be $\Delta t_{\rm THC0005}=3.55$ ms, $\Delta t_{\rm THC0032}=2.36$ ms, $\Delta t_{\rm THC0040}=3.12$ ms, respectively. In addition, we place physically informed priors on some parameters from existing phenomenological studies. 

\begin{table*}[t!]
\centering
\scalebox{1.0}{%
\begin{tabular}{rcccccc} \hline \hline   

Vector & \multicolumn{2}{ c }{THC0005} & \multicolumn{2}{ c }{THC0032}& \multicolumn{2}{ c }{THC0040}    \\ \hline  
\cline{2-7}

& Best Fit & \textbf{$\langle x\rangle \pm \sigma(x)$} & Best Fit & \textbf{$\langle x\rangle \pm \sigma(x)$} & Best Fit & \textbf{$\langle x\rangle \pm \sigma(x)$}   \\ \hline  

$k $ $[M_{\odot}/ms^{2}]$ & 2683.49& 2719.58 $\pm$ 54.37& 1658.17&  1648.53 $\pm$ 25.57& 2506.34&2551.38 $\pm$ 54.13\\ \hline  

$\alpha_2$ $[km/ms]$&103.18& 111.42  $\pm$12.19&2.11& 8.22 $\pm$ 8.01& 100.30&105.44 $\pm$ 10.68\\ \hline  

$\omega_0$ $[1/ms]$ &4.19& 4.25 $\pm$  0.06& 4.18& 4.12 $\pm$ 1.11& 4.24&  4.28 $\pm$ 1.93\\ \hline  

$r_0$   $[km]$ & 2.50& 2.59 $\pm$ 0.09& 2.94& 3.05 $\pm$ 0.19& 2.54&2.59 $\pm$ 0.73\\ \hline  

$l$ $[km]$& 6.75& 6.80  $\pm$0.16& 8.06& 8.26 $\pm$ 0.32& 7.07& 7.10 $\pm$ 1.11\\ \hline  

$\tilde{f}$ $[km/M_{\odot}]$ & 0.27& 0.24 $\pm$  0.03& 0.21& 0.20 $\pm$ 0.02& 0.33& 0.27 $\pm$ 0.07\\ \hline  

$m_1$ $[M_{\odot}]$ & 1.44& 1.41 $\pm$  0.03& 1.42& 1.42 $\pm$ 0.03& 1.48& 1.43 $\pm$ 0.11\\ \hline  

$m_2$ $[M_{\odot}]$& 1.18&  1.21 $\pm$ 0.03& 1.18& 1.18 $\pm$ 0.03& 1.26&  1.31$\pm$ 0.08\\ \hline\hline
\end{tabular}%
}
\caption{The values of the fitted 8-tuple fitting selected early post-merger GW waveforms in Fig.(\ref{fig:datas}) with no DM $f_\chi=\beta_{1,2}\equiv0$. The first column (vector) shows the 10-tuple quantities in  parameter space along with units. The Best Fit and   $\langle x\rangle\pm \sigma(x)$ columns are  obtained using the MCMC method. Details are described in Section \ref{MCMC setup}.}
\label{T:vectores}
\end{table*}

For example, for mass values, such as $m_1$ and $m_2$, stricter constraints are applied from the database instance itself, whereas for others, like $\alpha_2,\beta_n$, for which expected values are less clear, much broader priors are allowed.

Thus, we first identify a set of 10-tuples that fit the early post-merger window for each THC00XX simulation assuming no DM. We nevertheless allow at this point a residual non-zero viscous force $\propto \alpha_2$ related to the baryon content. The plot of polarizations with fixed parameters to the vectors of the best fit is shown in  Fig. (\ref{fig:Sol}).
More in detail we specify the best fit (as provided by emcee) and the median value along statistical sigma-deviation $\langle x\rangle \pm \sigma(x)$ for each fitted parameter. These values are summarized in the  Table (\ref{T:vectores}).


Elucidating the microscopic effect of the viscosity of matter in an extreme environment like this proves to be very challenging due to the assumed hybrid nature, ordinary and dark, of matter under study. We will assume that the self-interacting DM polluting the standard NS  matter may effectively impact emitted GW patterns while for self-interacting DM sector  we only have some astrophysical and cosmological hints obtained from phenomenology in large-scale structure  where some works \cite{Anand_2018} describe that CDM may acquire viscous character from decaying to relativistic particles or scattering ordinary matter. Recent simulations may also indicate additional effects \cite{GonzaloPhysRevLett.133.021401}. In our setting, we will consider the BNS merging event in a 
quasi-static phase approximation during which the density in the disk during the time window analyzed remains nearly constant so that we can characterize it by an effective hybrid kinetic viscosity value. 
\\

At this point, it is important to note that in our model dissipative forces are described by characteristic  $\alpha_2, \beta_n$ functions where we have made explicit their dependence 
in linear density. 

In general, a kinetic viscosity can be defined as $\alpha=\frac{\mu}{\rho}$ where $\mu$ is the dynamic viscosity and $\rho$ is the average volumetric mass density of the merging NS system. In order to convert quantities we write for our model 
\begin{equation}
l_c \rho_{\rm eff}\alpha_{\rm{eff}}=\alpha_2f_{\rm{ejec}}\frac{M_T}{R_d},
\end{equation}
where $R_d$ is the disk radius and $l_c$ represents a characteristic length of the system, physically motivated by the drifting objects, see below. Thus, if we define the density of the disk as that of a cylinder with a height $h< R_d$, we obtain
\begin{equation}
\rho_{\rm eff}=\frac{f_{\rm ejec}M_T}{\pi h R_d^{2}}.
\end{equation}
From this, it follows that 
\begin{equation}
\alpha_{\rm eff}=\pi\frac{h}{l_c}R_d \alpha_2.
\end{equation}
Given that we have taken an approximate radius $R_d \simeq 15.44$ km, we can place bounds on  the height of the matter disk as $2R_S <h<2R_d$ such that it is smaller than disk diameter and greater than twice the Schwarzschild radius, $R_S$, of the associated mass of the nuclei oscillating inside, we fix $h \sim 15$ km. Thus, we take the characteristic distance of the system such that it is on the order of twice the Schwarzschild radius, which holds for all the cases studied in this article as $l_c\lesssim 2G\bar{M}/c^2\sim 9$ km, $\bar{M} \equiv\left(m_1+m_2\right) / 2$. The conversion can be expressed as $\mu=\alpha_{\rm eff}\rho$. Using the values collected in Table \ref{T:vectores} and based on the above assumptions, we see the DM-baryon interaction is rather weak not only for data sets THC0005, THC0032 but also for  THC0040 where M0 neutrino scheme is used. In any case our setting must be refined to make any quantitative claim on the viscous effects from hadronic matter. 

\begin{table}
\centering
\scalebox{1.0}{%
\begin{tabular}{lccc}
\hline \hline DATA &$\alpha_{\rm eff}$~[$\rm cm^{2}/s$]& $\rho$ ~[$\rm g/cm^{3}$] & $\mu$~[$\rm g/cms$]  \\
\hline
THC0005 & $2.351\times10^{16}$ &  $0.274  \times 10^{14}$ & $0.644\times10^{30}$\\ 
THC0032 & $0.072\times10^{16}$ &  $0.220\times10^{14}$ &  $0.0158\times10^{30}$\\ 
THC0040 & $2.602\times10^{16}$ &  $0.345\times10^{14}$ & $0.898\times10^{30}$\\ 
\hline \hline
\end{tabular}}
\caption{ The effective kinetic viscosity ($\alpha_{\rm eff}$), density ($\rho$) and bulk viscosity ($\mu$) values for hybrid matter within the BNS merger disk, presented in the cgs system, corresponding to various datasets used in this work.}
\label{tab:den_vis}
\end{table}






\section{Dark Matter imprints on the spectral distribution}
\label{sec:PSD}

We now assess further qualitative changes from the GW patterns as obtained within our model for nearly symmetric BNS mergers. We do not attempt, as mentioned, precisely characterizing the impact of DM but rather spotting how viscous DM may influence the spectral distribution.  

On more technical grounds, one interesting property characterizing GW emission in BNS mergers appears to be the quasi-universal nature of their frequency spectrum for systems with $q=1$. However, some authors \cite{KiuchiPhysRevD.101.084006} claim in asymmetric systems tidal disruption of the NS cores ruins this picture. In their Fig. 11 and Fig. 12 in that work they show that the $f_1$ peak is less clearly detectable as BNS mass asymmetry $q$ grows. Their asymmetric cases departing from ($\eta=0.2500)$ or equivalently $q=1$ show the peak smears off. In our analysis we use $\eta=0.2494, 0.2485$ so we expect our qualitative conclusions still hold.  

}

\begin{figure*}[t]
\centering
\includegraphics[width=0.9\textwidth]{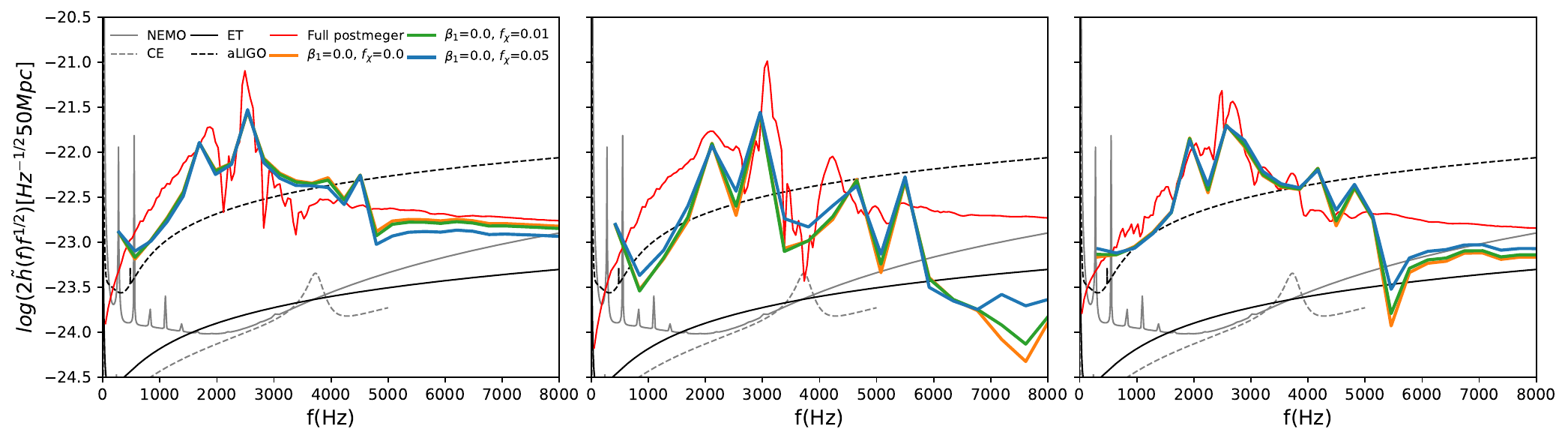}
\label{fig3:sub3}
\vspace{1mm} 

\includegraphics[width=0.9\textwidth]{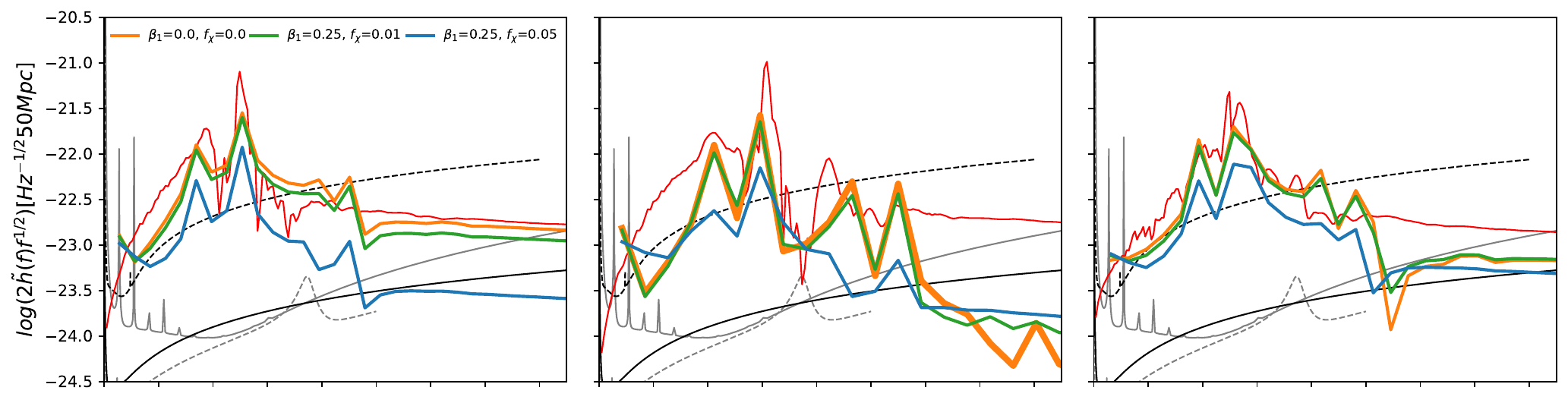}
\label{fig3:sub2}
\vspace{1mm} 

\includegraphics[width=0.91\textwidth]{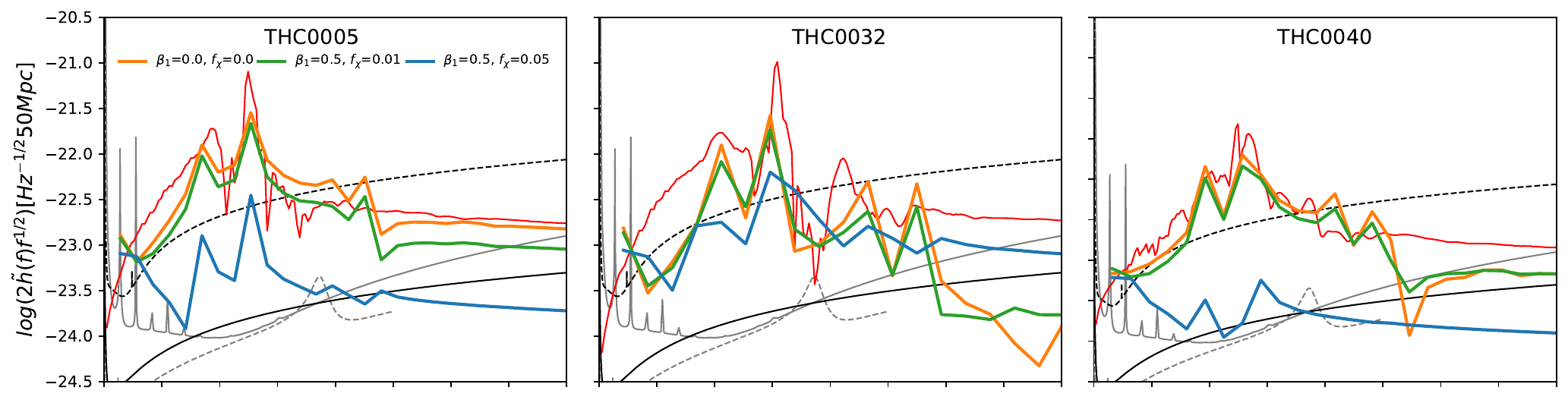}

\caption{Panel represents the spectral density $\sqrt{S_h(f)}=2\tilde{h}(f)f^{1/2}$ computed for the full post-merger data (thin red line) and for the early post-merger interval after the fitting procedure (thick colored lines) obtained from variation of DM parameters ($f_{\chi}$, $\beta_1$) with $\beta_2=0$. $\beta_1$ is expressed in $\rm 10^{16} (cm/s)^2$ units. We also plot sensitivity curves of the NEMO, CE, ET and aLIGO detectors (black and grey curves). Data sets THC0005, THC0032, and THC0040 appear in the left, center, right columns, respectively. The lower panel shows the case of $\beta_1=0$. In all cases $d=50$ Mpc.}
\label{fig:PSD_1}
\end{figure*}

We now set a distance $d=50$ Mpc and analyze the PSD, more precisely  taken as $2\tilde{h}(f) f^{1 / 2}$ where $\tilde{h}(f)$ is defined as the Fourier transform $\tilde{h}(f) \equiv \left|\tilde{h}_{+}(f)\right|$ and from the $h_+(t)$ polarization 
\begin{equation}\label{key}\tilde{h}_+(f)\equiv\int_{0}^{t_w}h_{+}(t)e^{-ift}dt,
\end{equation}
where $t_w$ is the early-postmerger window time $\Delta t$ that we defined for each signal, see Fig. (\ref{fig:Sol}). Note that from construction this will not match the full Fourier transform of the whole simulation data even in absence of DM but we take this as a benchmark to compare our findings. 
The transform is taken over this polarization mode due to the approximation to the principal mode $h_{+}\simeq h^{22}$.

Now we can calculate the logarithm of the spectral density $\log_{10} \left(2 \tilde{h}(f) f^{1 / 2}\right)\left[\rm Hz^{-1 / 2}\ 50 \mathrm{Mpc}\right]$ and study, first, the benchmark results from DM free selected CoRe database simulations and then proceed to see the impact of the (viscous) dark component. 

  
\begin{figure*}[t]
\centering
\includegraphics[width=0.9\textwidth]{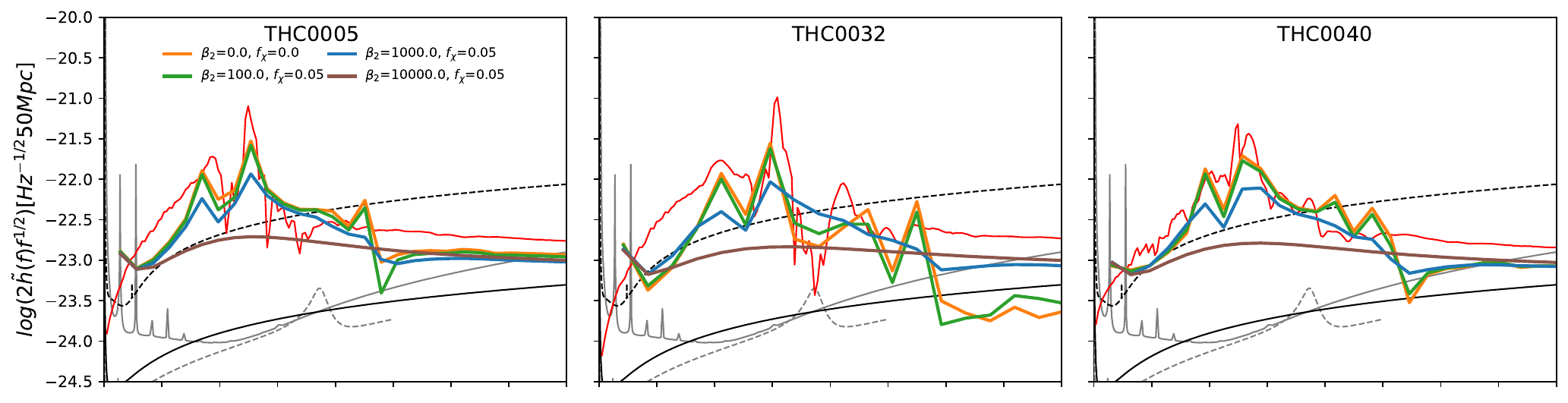}
\label{fig:sub2}
\vspace{0.1mm}

\includegraphics[width=0.9\textwidth]{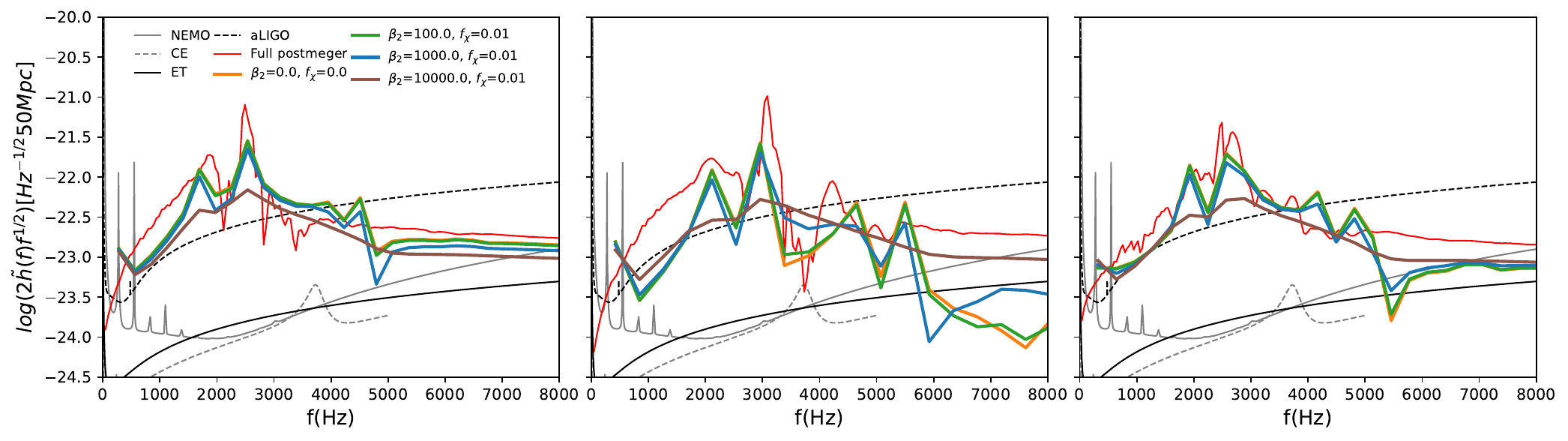}
\label{fig:sub1}

\caption{Panel represents the spectral density $\sqrt{S_h(f)}=2\tilde{h}(f)f^{1/2}$ computed for the full post-merger data (thin red line) and for the early post-merger interval after the fitting procedure (thick colored lines) obtained from variation of DM parameters ($f_{\chi}$, $\beta_2$) with $\beta_1=0$. We also plot sensitivity curves of the NEMO, CE, ET and aLIGO detectors (black and grey curves). Data sets THC0005, THC0032, and THC0040 appear in the left, center, right columns, respectively. Distance is fixed to $d=50$ Mpc.}
\label{fig:PSD_2}
\end{figure*}

In Fig. (\ref{fig:PSD_1}) we plot our results for the three data sets THC0005, THC0032, and THC0040 in the left, center, and right columns, respectively. We depict with a thin red line the full post-merger stage of the data and with thick orange, green, and blue lines the early post-merger interval $\Delta t$ associated with each dataset utilized. We set $\beta_2=0$ in this panel for all studied cases. The orange line shows the DM free case, while the orange and blue set $f_\chi=1\%$ and $f_\chi=5\%$ DM mass fraction, respectively. For the upper panel $\beta_1=0.5\times10^{16}$ $\rm (cm/s)^2$, middle $\beta_1=0.25\times10^{16}$ $\rm (cm/s)^2$ and lower $\beta_1=0$. In order to see how feasible a future detection may be we superimpose the sensitivity curves of the Advanced LIGO, ET, NEMO and Cosmic Explorer detectors. We obtain $f_{\rm max}$=2.3426, 2.9608 and 2.6667  kHz for THC0005, THC0032,THC0040 cases, respectively. 
For our set of EoS the nuclear matter incompressibility $K\in[242.7,245.5]$ MeV at saturation density, thus to this respect we are showing cases with a small stiffness spread. We nevertheless see by simple inspection that small compactness stars are more sensitive to a $f_\chi\neq 0$ fraction and thus more likely to show deviations of the DM free $f_{\rm max}$ value \cite{Mariani_2023}.

Upon analyzing the frequency spectrum obtained from our model, it becomes evident that for the nearly asymmetric cases we consider this model reasonably provides the principal peak $f_{\rm max}$, which characterizes the post-merger signal, across the entire temporal interval of the simulated signal. 

Further comparison shows the model captures a secondary, lower frequency peak at low frequencies associated with a non-linear coupling between the quadrupolar oscillation modes and the quasi-radial modes.
It should be noted, in addition, that the primary contributions belong to kHz frequencies, where the peaks appear to originate almost entirely during the early post-merger stage, as illustrated in \cite{takamiPhysRevD.91.064001} for the peak at lower frequencies. Similarly, concerning the largest amplitude peak, it has been observed to grow rapidly in the early post-merger stage and then remain almost unchanged throughout the remainder of the post-merger period \cite{soultanis2022analytic}.

In the following, we discuss the obtained results for the PSD.  We first consider that the environmental DM causes dissipation in the system that is constant and proportional to linear DM density. This behavior is sometimes referred to in the literature as {\it Coulomb damping} \cite{fay2012coulomb}. In this case, there is only a non-zero $\beta_n$ i.e. $\beta_1\neq0$ in Section \ref{sec:model} and we fix $\beta_2=0$. For decreasing values of $\beta_1$ characterizing DM  behavior we show  (left, center and right panels) the PSD for the THC0005, THC0032, and THC0040 cases (thick colored lines) considered in Fig. (\ref{fig:PSD_1}). We set $d=50$ Mpc. The amplitude of the frequency peaks exhibits more pronounced deviations from the DM-free case ($f_\chi=\beta_n=0$).  This is especially relevant in the right end of the $f_\chi \in [0.01,0.05]$ interval.
Importantly, the decrease in the amplitudes in $\sqrt{S_h(f)}=2\tilde{h}(f)f^{\frac{1}{2}}$ from DM dissipation may result in the post-merger region of the event no longer being resolved by the detector at that distance. Consequently, events could be detected at distances where the inspiral phase is well-determined, but not that of the post-merger phase. In other words, detections of BNS  mergers in environments containing modest quantities of viscous DM might exhibit GW emission where the typical post-merger early-phase behavior is not discernible, resembling, instead, a rapid decay toward a BH, for example.
As shown in the lower panels, when the fraction of DM remains below the fiducial $\sim 5\%$ threshold and it is assumed a non-viscous behavior, the spectral density curve exhibits only marginal deviations from the DM free pattern. This observation suggests that the introduction of a limited amount of non-viscous matter does not significantly disrupt the expected spectral characteristics. 

In Fig.(\ref{fig:PSD_2}) we analyze cases with $\beta_1\equiv 0$ to examine the frequency spectrum for different values of $\beta_2\neq0$, using the same mass fractions $f_\chi=0, 0.01, 0.05$ as in the previous Fig. (\ref{fig:PSD_1}). We can observe that the characteristic peaks of the PSD curve disappear. For the case  with $f_\chi=0.05$ and $\beta_2=10^{4}$ (in $\rm 10^8\, cm/s$ units), the signal vanished entirely, in contrast with the behavior obtained for DM with Coulomb-type damping, where the signal peak amplitude decreases while maintaining its characteristic shape.



In order to discuss the prospects of detectability we proceed to examine the damping or attenuation in the maximum peak amplitude \( f_{\text{max}} \)  in the PSD due to the existence of a DM polluted environment. For this we calculate the effective signal-to-noise ratio (eSNR) as done in \cite{takamiPhysRevD.91.064001}. In our case this could be defined as 

\begin{equation}
\rm eSNR \equiv \frac{2\tilde{h}(f_{\text{max}})\sqrt{f_{\text{max}}}}{S_n(f_{\text{max}})}
\end{equation}

where \( S_n(f_{\text{max}}) \) is the value of the detector's sensitivity curve at the frequency where maximum peak amplitude of the GW signal appears. 

The eSNR, as a function of the DM phase space $(\beta_1, f_{\chi})$, constitutes a useful and straightforward tool that allows sizing the influence of DM on the attenuation of peak amplitudes and the rest of the modes. Note that, in principle, this quantity is frequency dependent, eSNR(f).
Considering the lower bound given by the sensitivity curve of aLIGO (indicated by the black dashed lines in Fig. (\ref{fig:PSD_1}), we observe  in Fig. (\ref{fig:THC32var}) the eSNR isocurves dependence on   ($\beta_1, f_\chi$) for the THC0032 case (bottom) and in more detail the PSD (top) for selected $\beta_1=0.33,0.37,0.4,0.42, 0.5$ in $\rm 10^{16} (cm/s)^2$ units and $f_\chi=0.045$ values.
Since $\beta_2$ largely impacts the damping behavior we choose for the sake of illustration, instead, the $\beta_1$  damping. For all cases, we assumed GW detections with optimal orientation at 50 Mpc. At this point we remark that the spectral analysis we perform is conducted within a $\Delta t$ temporal window  and consequently, the depicted PSD reflects only the frequencies that influence the signal's power within this interval.
\begin{figure}[!h]
\centering
\includegraphics[width=0.5\textwidth]{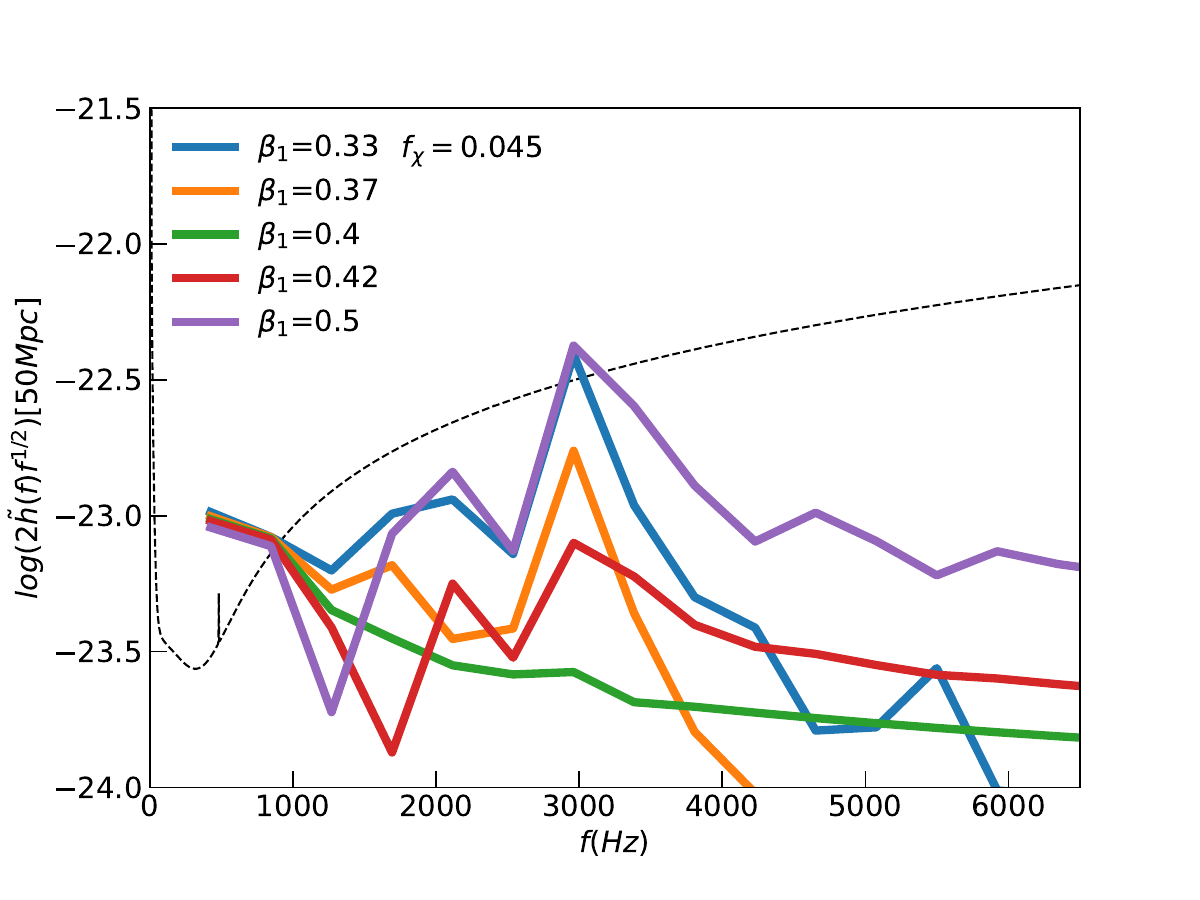}
\label{fig:THC32PSD}


\includegraphics[width=0.55\textwidth]{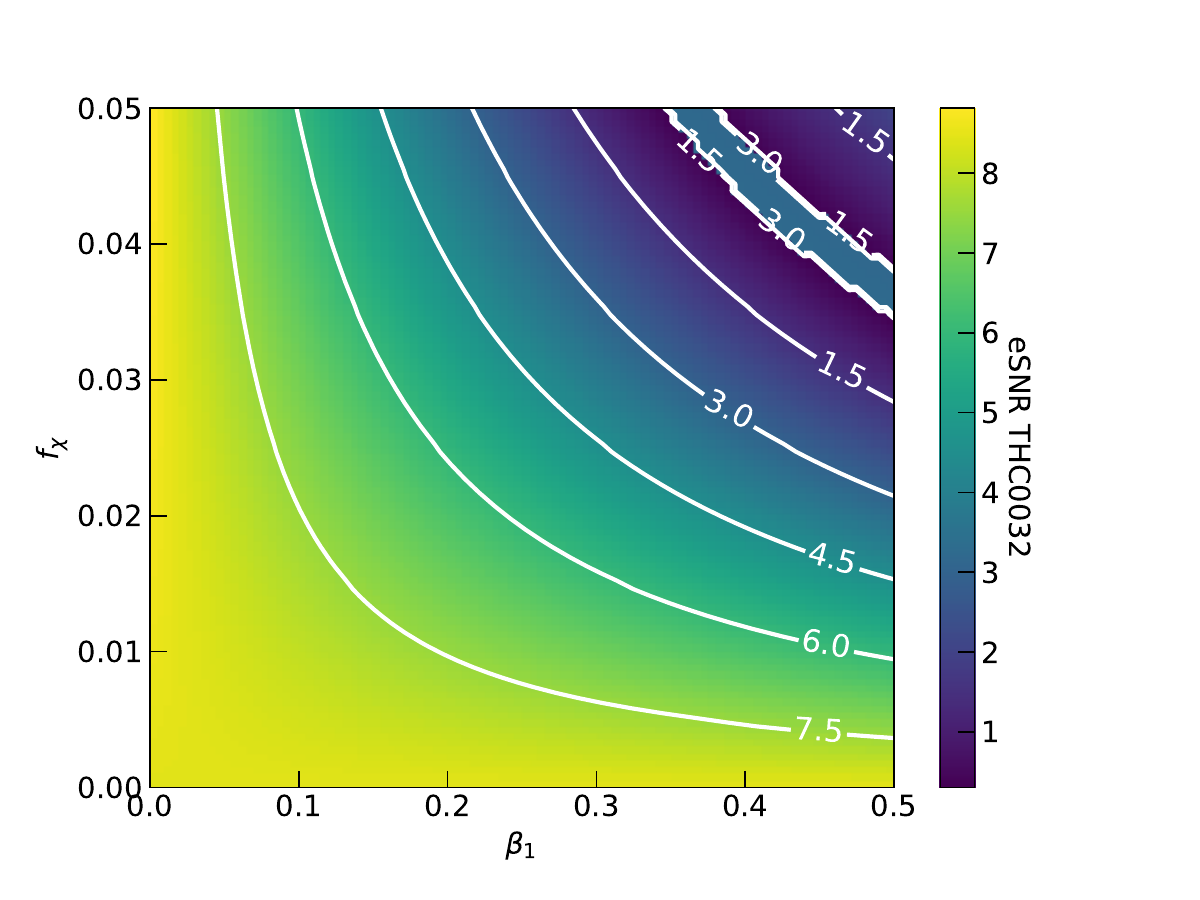}
\label{fig:THC32eSNR}

\caption{eSNR as a function of the DM parameters ($\beta_1, f_\chi$) for THC0032 (bottom) and PSD (top) with respect to the aLIGO (black dashed line in Fig. \ref{fig:PSD_1}) for $\beta_1=0.33,0.37,0.4,0.42, 0.5$ in $\rm 10^{16} (cm/s)^2$ units and $f_\chi=0.045$ where a non-monotonic eSNR change appears. For all cases, we assumed optimal orientation at $d=50$ Mpc.}
\label{fig:THC32var}
\end{figure}


We rely on previous results, which suggest that the main peak is predominantly associated with the early post-merger phase. During this interval, kHz frequencies are more prominent. However, when DM is introduced, the main peak amplitude decreases, reducing the contribution of these frequencies to the signal. From inspection of the eSNR in the bottom panel we see that there is an enhancement, breaking the attenuation tendency,  of the peak amplitude for a band (upper  right corner). This is due to the fact that in this band the peak amplitude shifts towards lower frequency $\sim 500$ Hz (thus reversing the eSNR ratio decrease)  and the original amplitude around $\sim3$ kHz falls within aLIGO sensitivity. The detail of this behavior is illustrated in the top panel where for $\beta_1\in[0.33,0.5]$ and $f_\chi=0.045$ the monotony of the main amplitude ratio changes.



\section{Conclusions}
\label{sec:Conclusions}
Taking an effective Lagrangian model partly inspired by \cite{takamiPhysRevD.91.064001,ELLIS2018607} we qualitatively explore the dynamics of nearly symmetric binary NS mergers in the presence of a DM-polluted environment. As a benchmark, this model allows a reasonable reproduction of the emitted GW signal in a time window for selected early post-merger instances of the CoRe simulation database with ordinary matter. 
We use MCMC techniques fitting a n-tuple parameter for each selected BNS configuration. 
We also obtained the resulting frequency spectrum and confirmed that the characteristic large amplitude mode, $f_{\rm max}$  can be recovered in this scenario where $q\sim 1$. 
When the initial (DM free) model is supplemented in a minimal procedure with a non-zero viscous mass fraction of DM in the BNS environment, the qualitative behavior of the GW changes exhibiting a pronounced damping. By analyzing the shape of the frequency spectrum, particularly its characteristic large amplitude peak, we tested the implications of adding up to realistic $5\%$ of matter from the environment. We  analyzed cases where DM matter was present but not inducing dissipative effects or it produced dissipation instead. 
For the inviscid matter, we found negligible changes in the PSD up to $5\%$ DM analyzed. For Coulomb/viscous  damping, we found that the amplitude of the main peak decreased appreciably, not resulting shifted from the $\sim3$ kHz value except for a restricted DM parameter region. However frequency dependent DM-induced attenuation does not affect low frequency models as much. This is suggesting that matter with this type of behavior may lead to detections where this peak is not resolved by interferometers and could mimic instead a prompt collapse to a more compact object. When considering the effective SNR we find aLIGO yields a lower bound and could detect the possible effects of damping in maximum amplitude mode yielding a non-monotonic tendency for a subset of DM parameters. There is thus a possible interplay among (ordinary) matter effects and those of viscous DM that should not be overlooked as pointed out by recent works \cite{GonzaloPhysRevLett.133.021401}. Further work is necessary to investigate these aspects and elucidate the importance of environmental effects and its imprint in waveforms.

\section*{Acknowledgements}
We acknowledge useful discussion from D. Radice, A. Perego, I. de Martino, A. de la Cruz-Dombriz. R. Della Monica and P. Char. We also acknowledge support from Spanish Ministry of Science PID2022-137887NB-10 and RED2022-134411-T projects, Junta de Castilla y León  projects SA101P24, SA091P24 and  European Union's HORIZON MSCA-2022-PF-01-01, project ProMatEx-NS. This publication is based upon work from the COST Action “COSMIC WISPers” (CA21106) supported by COST (European Cooperation in Science and Technology). D. Suarez-Fontanella gratefully acknowledges the financial support provided by the Banco Santander International Doctoral Scholarship. D. Barba-González acknowledges support from a Ph.D. Fellowship funded by Consejería de Educación de la Junta de Castilla y León and European Social Fund.We acknowledge use of Spanish RES resources.

\bibliographystyle{apsrev} 

\bibliography{refs}

\begin{thebibliography}{58}
\expandafter\ifx\csname natexlab\endcsname\relax\def\natexlab#1{#1}\fi
\expandafter\ifx\csname bibnamefont\endcsname\relax
  \def\bibnamefont#1{#1}\fi
\expandafter\ifx\csname bibfnamefont\endcsname\relax
  \def\bibfnamefont#1{#1}\fi
\expandafter\ifx\csname citenamefont\endcsname\relax
  \def\citenamefont#1{#1}\fi
\expandafter\ifx\csname url\endcsname\relax
  \def\url#1{\texttt{#1}}\fi
\expandafter\ifx\csname urlprefix\endcsname\relax\def\urlprefix{URL }\fi
\providecommand{\bibinfo}[2]{#2}
\providecommand{\eprint}[2][]{\url{#2}}

\bibitem[{\citenamefont{Takami et~al.}(2015)\citenamefont{Takami, Rezzolla, and
  Baiotti}}]{takamiPhysRevD.91.064001}
\bibinfo{author}{\bibfnamefont{K.}~\bibnamefont{Takami}},
  \bibinfo{author}{\bibfnamefont{L.}~\bibnamefont{Rezzolla}}, \bibnamefont{and}
  \bibinfo{author}{\bibfnamefont{L.}~\bibnamefont{Baiotti}},
  \bibinfo{journal}{Phys. Rev. D} \textbf{\bibinfo{volume}{91}},
  \bibinfo{pages}{064001} (\bibinfo{year}{2015}),
  \urlprefix\url{https://link.aps.org/doi/10.1103/PhysRevD.91.064001}.

\bibitem[{\citenamefont{Arbey and Mahmoudi}(2021)}]{Arbey:2021gdg}
\bibinfo{author}{\bibfnamefont{A.}~\bibnamefont{Arbey}} \bibnamefont{and}
  \bibinfo{author}{\bibfnamefont{F.}~\bibnamefont{Mahmoudi}},
  \bibinfo{journal}{Prog. Part. Nucl. Phys.} \textbf{\bibinfo{volume}{119}},
  \bibinfo{pages}{103865} (\bibinfo{year}{2021}), \eprint{2104.11488}.

\bibitem[{\citenamefont{Marsh et~al.}(2024)\citenamefont{Marsh, Ellis, and
  Mehta}}]{Marsh:2024ury}
\bibinfo{author}{\bibfnamefont{D.~J.~E.} \bibnamefont{Marsh}},
  \bibinfo{author}{\bibfnamefont{D.}~\bibnamefont{Ellis}}, \bibnamefont{and}
  \bibinfo{author}{\bibfnamefont{V.~M.} \bibnamefont{Mehta}},
  \emph{\bibinfo{title}{{Dark Matter: Evidence, Theory, and Constraints}}}
  (\bibinfo{publisher}{Princeton University Press}, \bibinfo{year}{2024}), ISBN
  \bibinfo{isbn}{978-0-691-24952-0}.

\bibitem[{\citenamefont{Mariani et~al.}(2023)\citenamefont{Mariani, Albertus,
  Alessandroni, Orsaria, P\'erez-García, and Ranea-Sandoval}}]{Mariani_2023}
\bibinfo{author}{\bibfnamefont{M.}~\bibnamefont{Mariani}},
  \bibinfo{author}{\bibfnamefont{C.}~\bibnamefont{Albertus}},
  \bibinfo{author}{\bibfnamefont{M.~d.~R.} \bibnamefont{Alessandroni}},
  \bibinfo{author}{\bibfnamefont{M.~G.} \bibnamefont{Orsaria}},
  \bibinfo{author}{\bibfnamefont{M.~A.} \bibnamefont{P\'erez-García}},
  \bibnamefont{and} \bibinfo{author}{\bibfnamefont{I.~F.}
  \bibnamefont{Ranea-Sandoval}}, \bibinfo{journal}{Monthly Notices of the Royal
  Astronomical Society} \textbf{\bibinfo{volume}{527}},
  \bibinfo{pages}{6795–6806} (\bibinfo{year}{2023}), ISSN
  \bibinfo{issn}{1365-2966},
  \urlprefix\url{http://dx.doi.org/10.1093/mnras/stad3658}.

\bibitem[{\citenamefont{{Rafiei Karkevandi} et~al.}(2024)\citenamefont{{Rafiei
  Karkevandi}, {Shahrbaf}, {Shakeri}, and {Typel}}}]{2024Parti...7..201R}
\bibinfo{author}{\bibfnamefont{D.}~\bibnamefont{{Rafiei Karkevandi}}},
  \bibinfo{author}{\bibfnamefont{M.}~\bibnamefont{{Shahrbaf}}},
  \bibinfo{author}{\bibfnamefont{S.}~\bibnamefont{{Shakeri}}},
  \bibnamefont{and} \bibinfo{author}{\bibfnamefont{S.}~\bibnamefont{{Typel}}},
  \bibinfo{journal}{Particles} \textbf{\bibinfo{volume}{7}},
  \bibinfo{pages}{201} (\bibinfo{year}{2024}), \eprint{2402.18696}.

\bibitem[{\citenamefont{Oks}(2021)}]{Oks_2021}
\bibinfo{author}{\bibfnamefont{E.}~\bibnamefont{Oks}}, \bibinfo{journal}{New
  Astronomy Reviews} \textbf{\bibinfo{volume}{93}}, \bibinfo{pages}{101632}
  (\bibinfo{year}{2021}), ISSN \bibinfo{issn}{1387-6473},
  \urlprefix\url{http://dx.doi.org/10.1016/j.newar.2021.101632}.

\bibitem[{\citenamefont{Zwicky}(1933)}]{zwicky1933rotverschiebung}
\bibinfo{author}{\bibfnamefont{F.}~\bibnamefont{Zwicky}},
  \bibinfo{journal}{Helvetica Physica Acta, Vol. 6, p. 110-127}
  \textbf{\bibinfo{volume}{6}}, \bibinfo{pages}{110} (\bibinfo{year}{1933}).

\bibitem[{\citenamefont{Bertone et~al.}(2020)\citenamefont{Bertone, Croon,
  Amin, Boddy, Kavanagh, Mack, Natarajan, Opferkuch, Schutz, Takhistov
  et~al.}}]{Bertone_2020}
\bibinfo{author}{\bibfnamefont{G.}~\bibnamefont{Bertone}},
  \bibinfo{author}{\bibfnamefont{D.}~\bibnamefont{Croon}},
  \bibinfo{author}{\bibfnamefont{M.}~\bibnamefont{Amin}},
  \bibinfo{author}{\bibfnamefont{K.~K.} \bibnamefont{Boddy}},
  \bibinfo{author}{\bibfnamefont{B.}~\bibnamefont{Kavanagh}},
  \bibinfo{author}{\bibfnamefont{K.~J.} \bibnamefont{Mack}},
  \bibinfo{author}{\bibfnamefont{P.}~\bibnamefont{Natarajan}},
  \bibinfo{author}{\bibfnamefont{T.}~\bibnamefont{Opferkuch}},
  \bibinfo{author}{\bibfnamefont{K.}~\bibnamefont{Schutz}},
  \bibinfo{author}{\bibfnamefont{V.}~\bibnamefont{Takhistov}},
  \bibnamefont{et~al.}, \bibinfo{journal}{SciPost Physics Core}
  \textbf{\bibinfo{volume}{3}} (\bibinfo{year}{2020}), ISSN
  \bibinfo{issn}{2666-9366},
  \urlprefix\url{http://dx.doi.org/10.21468/SciPostPhysCore.3.2.007}.

\bibitem[{\citenamefont{Abbott et~al.}(2016)\citenamefont{Abbott, Abbott,
  Abbott, Abernathy, Acernese, Ackley, Adams, Adams, Addesso, Adhikari
  et~al.}}]{ligobbh}
\bibinfo{author}{\bibfnamefont{B.~P.} \bibnamefont{Abbott}},
  \bibinfo{author}{\bibfnamefont{R.}~\bibnamefont{Abbott}},
  \bibinfo{author}{\bibfnamefont{T.~D.} \bibnamefont{Abbott}},
  \bibinfo{author}{\bibfnamefont{M.~R.} \bibnamefont{Abernathy}},
  \bibinfo{author}{\bibfnamefont{F.}~\bibnamefont{Acernese}},
  \bibinfo{author}{\bibfnamefont{K.}~\bibnamefont{Ackley}},
  \bibinfo{author}{\bibfnamefont{C.}~\bibnamefont{Adams}},
  \bibinfo{author}{\bibfnamefont{T.}~\bibnamefont{Adams}},
  \bibinfo{author}{\bibfnamefont{P.}~\bibnamefont{Addesso}},
  \bibinfo{author}{\bibfnamefont{R.~X.} \bibnamefont{Adhikari}},
  \bibnamefont{et~al.} (\bibinfo{collaboration}{LIGO Scientific Collaboration
  and Virgo Collaboration}), \bibinfo{journal}{Phys. Rev. Lett.}
  \textbf{\bibinfo{volume}{116}}, \bibinfo{pages}{131103}
  (\bibinfo{year}{2016}),
  \urlprefix\url{https://link.aps.org/doi/10.1103/PhysRevLett.116.131103}.

\bibitem[{\citenamefont{Abbott et~al.}(2023)\citenamefont{Abbott, Abbott,
  Acernese, Ackley, Adams, Adhikari, Adhikari, Adya, Affeldt, Zhao
  et~al.}}]{rate}
\bibinfo{author}{\bibfnamefont{R.}~\bibnamefont{Abbott}},
  \bibinfo{author}{\bibfnamefont{T.~D.} \bibnamefont{Abbott}},
  \bibinfo{author}{\bibfnamefont{F.}~\bibnamefont{Acernese}},
  \bibinfo{author}{\bibfnamefont{K.}~\bibnamefont{Ackley}},
  \bibinfo{author}{\bibfnamefont{C.}~\bibnamefont{Adams}},
  \bibinfo{author}{\bibfnamefont{N.}~\bibnamefont{Adhikari}},
  \bibinfo{author}{\bibfnamefont{R.~X.} \bibnamefont{Adhikari}},
  \bibinfo{author}{\bibfnamefont{V.~B.} \bibnamefont{Adya}},
  \bibinfo{author}{\bibfnamefont{Y.}~\bibnamefont{Affeldt},
  \bibfnamefont{C.~anang}},
  \bibinfo{author}{\bibfnamefont{C.}~\bibnamefont{Zhao}}, \bibnamefont{et~al.}
  (\bibinfo{collaboration}{LIGO Scientific Collaboration, Virgo Collaboration,
  and KAGRA Collaboration}), \bibinfo{journal}{Phys. Rev. X}
  \textbf{\bibinfo{volume}{13}}, \bibinfo{pages}{011048}
  (\bibinfo{year}{2023}),
  \urlprefix\url{https://link.aps.org/doi/10.1103/PhysRevX.13.011048}.

\bibitem[{\citenamefont{Nitz et~al.}(2023)\citenamefont{Nitz, Kumar, Wang,
  Kastha, Wu, Schäfer, Dhurkunde, and Capano}}]{Nitz_2023}
\bibinfo{author}{\bibfnamefont{A.~H.} \bibnamefont{Nitz}},
  \bibinfo{author}{\bibfnamefont{S.}~\bibnamefont{Kumar}},
  \bibinfo{author}{\bibfnamefont{Y.-F.} \bibnamefont{Wang}},
  \bibinfo{author}{\bibfnamefont{S.}~\bibnamefont{Kastha}},
  \bibinfo{author}{\bibfnamefont{S.}~\bibnamefont{Wu}},
  \bibinfo{author}{\bibfnamefont{M.}~\bibnamefont{Schäfer}},
  \bibinfo{author}{\bibfnamefont{R.}~\bibnamefont{Dhurkunde}},
  \bibnamefont{and} \bibinfo{author}{\bibfnamefont{C.~D.}
  \bibnamefont{Capano}}, \bibinfo{journal}{The Astrophysical Journal}
  \textbf{\bibinfo{volume}{946}}, \bibinfo{pages}{59} (\bibinfo{year}{2023}),
  \urlprefix\url{https://dx.doi.org/10.3847/1538-4357/aca591}.

\bibitem[{\citenamefont{Evans et~al.}(2023)\citenamefont{Evans, Corsi, Afle,
  Ananyeva, Arun, Ballmer, Bandopadhyay, Barsotti, Baryakhtar, Berger
  et~al.}}]{evans2023cosmicexplorersubmissionnsf}
\bibinfo{author}{\bibfnamefont{M.}~\bibnamefont{Evans}},
  \bibinfo{author}{\bibfnamefont{A.}~\bibnamefont{Corsi}},
  \bibinfo{author}{\bibfnamefont{C.}~\bibnamefont{Afle}},
  \bibinfo{author}{\bibfnamefont{A.}~\bibnamefont{Ananyeva}},
  \bibinfo{author}{\bibfnamefont{K.~G.} \bibnamefont{Arun}},
  \bibinfo{author}{\bibfnamefont{S.}~\bibnamefont{Ballmer}},
  \bibinfo{author}{\bibfnamefont{A.}~\bibnamefont{Bandopadhyay}},
  \bibinfo{author}{\bibfnamefont{L.}~\bibnamefont{Barsotti}},
  \bibinfo{author}{\bibfnamefont{M.}~\bibnamefont{Baryakhtar}},
  \bibinfo{author}{\bibfnamefont{E.}~\bibnamefont{Berger}},
  \bibnamefont{et~al.}, \emph{\bibinfo{title}{Cosmic explorer: A submission to
  the nsf mpsac nggw subcommittee}} (\bibinfo{year}{2023}),
  \eprint{2306.13745}, \urlprefix\url{https://arxiv.org/abs/2306.13745}.

\bibitem[{\citenamefont{Branchesi et~al.}(2023)\citenamefont{Branchesi,
  Maggiore, Alonso, Badger, Banerjee, Beirnaert, Belgacem, Bhagwat, Boileau,
  Borhanian et~al.}}]{Branchesi_2023}
\bibinfo{author}{\bibfnamefont{M.}~\bibnamefont{Branchesi}},
  \bibinfo{author}{\bibfnamefont{M.}~\bibnamefont{Maggiore}},
  \bibinfo{author}{\bibfnamefont{D.}~\bibnamefont{Alonso}},
  \bibinfo{author}{\bibfnamefont{C.}~\bibnamefont{Badger}},
  \bibinfo{author}{\bibfnamefont{B.}~\bibnamefont{Banerjee}},
  \bibinfo{author}{\bibfnamefont{F.}~\bibnamefont{Beirnaert}},
  \bibinfo{author}{\bibfnamefont{E.}~\bibnamefont{Belgacem}},
  \bibinfo{author}{\bibfnamefont{S.}~\bibnamefont{Bhagwat}},
  \bibinfo{author}{\bibfnamefont{G.}~\bibnamefont{Boileau}},
  \bibinfo{author}{\bibfnamefont{S.}~\bibnamefont{Borhanian}},
  \bibnamefont{et~al.}, \bibinfo{journal}{Journal of Cosmology and
  Astroparticle Physics} \textbf{\bibinfo{volume}{2023}}, \bibinfo{pages}{068}
  (\bibinfo{year}{2023}), ISSN \bibinfo{issn}{1475-7516},
  \urlprefix\url{http://dx.doi.org/10.1088/1475-7516/2023/07/068}.

\bibitem[{\citenamefont{Ackley et~al.}(2020)\citenamefont{Ackley, Adya,
  Agrawal, Altin, Ashton, Bailes, Baltinas, Barbuio, Beniwal, Blair
  et~al.}}]{nemoAckley_2020}
\bibinfo{author}{\bibfnamefont{K.}~\bibnamefont{Ackley}},
  \bibinfo{author}{\bibfnamefont{V.~B.} \bibnamefont{Adya}},
  \bibinfo{author}{\bibfnamefont{P.}~\bibnamefont{Agrawal}},
  \bibinfo{author}{\bibfnamefont{P.}~\bibnamefont{Altin}},
  \bibinfo{author}{\bibfnamefont{G.}~\bibnamefont{Ashton}},
  \bibinfo{author}{\bibfnamefont{M.}~\bibnamefont{Bailes}},
  \bibinfo{author}{\bibfnamefont{E.}~\bibnamefont{Baltinas}},
  \bibinfo{author}{\bibfnamefont{A.}~\bibnamefont{Barbuio}},
  \bibinfo{author}{\bibfnamefont{D.}~\bibnamefont{Beniwal}},
  \bibinfo{author}{\bibfnamefont{C.}~\bibnamefont{Blair}},
  \bibnamefont{et~al.}, \bibinfo{journal}{Publications of the Astronomical
  Society of Australia} \textbf{\bibinfo{volume}{37}} (\bibinfo{year}{2020}),
  ISSN \bibinfo{issn}{1448-6083},
  \urlprefix\url{http://dx.doi.org/10.1017/pasa.2020.39}.

\bibitem[{\citenamefont{Amaro-Seoane et~al.}(2017)\citenamefont{Amaro-Seoane,
  Audley, Babak, Baker, Barausse, Bender, Berti, Binetruy, Born, Bortoluzzi
  et~al.}}]{amaroseoane2017laserinterferometerspaceantenna}
\bibinfo{author}{\bibfnamefont{P.}~\bibnamefont{Amaro-Seoane}},
  \bibinfo{author}{\bibfnamefont{H.}~\bibnamefont{Audley}},
  \bibinfo{author}{\bibfnamefont{S.}~\bibnamefont{Babak}},
  \bibinfo{author}{\bibfnamefont{J.}~\bibnamefont{Baker}},
  \bibinfo{author}{\bibfnamefont{E.}~\bibnamefont{Barausse}},
  \bibinfo{author}{\bibfnamefont{P.}~\bibnamefont{Bender}},
  \bibinfo{author}{\bibfnamefont{E.}~\bibnamefont{Berti}},
  \bibinfo{author}{\bibfnamefont{P.}~\bibnamefont{Binetruy}},
  \bibinfo{author}{\bibfnamefont{M.}~\bibnamefont{Born}},
  \bibinfo{author}{\bibfnamefont{D.}~\bibnamefont{Bortoluzzi}},
  \bibnamefont{et~al.}, \emph{\bibinfo{title}{Laser interferometer space
  antenna}} (\bibinfo{year}{2017}), \eprint{1702.00786},
  \urlprefix\url{https://arxiv.org/abs/1702.00786}.

\bibitem[{\citenamefont{Ellis et~al.}(2018)\citenamefont{Ellis, Hektor, Hütsi,
  Kannike, Marzola, Raidal, and Vaskonen}}]{ELLIS2018607}
\bibinfo{author}{\bibfnamefont{J.}~\bibnamefont{Ellis}},
  \bibinfo{author}{\bibfnamefont{A.}~\bibnamefont{Hektor}},
  \bibinfo{author}{\bibfnamefont{G.}~\bibnamefont{Hütsi}},
  \bibinfo{author}{\bibfnamefont{K.}~\bibnamefont{Kannike}},
  \bibinfo{author}{\bibfnamefont{L.}~\bibnamefont{Marzola}},
  \bibinfo{author}{\bibfnamefont{M.}~\bibnamefont{Raidal}}, \bibnamefont{and}
  \bibinfo{author}{\bibfnamefont{V.}~\bibnamefont{Vaskonen}},
  \bibinfo{journal}{Physics Letters B} \textbf{\bibinfo{volume}{781}},
  \bibinfo{pages}{607} (\bibinfo{year}{2018}), ISSN \bibinfo{issn}{0370-2693},
  \urlprefix\url{https://www.sciencedirect.com/science/article/pii/S0370269318303423}.

\bibitem[{\citenamefont{Ashton et~al.}(2019)\citenamefont{Ashton, Hübner,
  Lasky, Talbot, Ackley, Biscoveanu, Chu, Divakarla, Easter, Goncharov
  et~al.}}]{Ashton_2019}
\bibinfo{author}{\bibfnamefont{G.}~\bibnamefont{Ashton}},
  \bibinfo{author}{\bibfnamefont{M.}~\bibnamefont{Hübner}},
  \bibinfo{author}{\bibfnamefont{P.~D.} \bibnamefont{Lasky}},
  \bibinfo{author}{\bibfnamefont{C.}~\bibnamefont{Talbot}},
  \bibinfo{author}{\bibfnamefont{K.}~\bibnamefont{Ackley}},
  \bibinfo{author}{\bibfnamefont{S.}~\bibnamefont{Biscoveanu}},
  \bibinfo{author}{\bibfnamefont{Q.}~\bibnamefont{Chu}},
  \bibinfo{author}{\bibfnamefont{A.}~\bibnamefont{Divakarla}},
  \bibinfo{author}{\bibfnamefont{P.~J.} \bibnamefont{Easter}},
  \bibinfo{author}{\bibfnamefont{B.}~\bibnamefont{Goncharov}},
  \bibnamefont{et~al.}, \bibinfo{journal}{The Astrophysical Journal Supplement
  Series} \textbf{\bibinfo{volume}{241}}, \bibinfo{pages}{27}
  (\bibinfo{year}{2019}),
  \urlprefix\url{https://dx.doi.org/10.3847/1538-4365/ab06fc}.

\bibitem[{\citenamefont{{Garani} et~al.}(2023)\citenamefont{{Garani}, {Raj},
  and {Reynoso-Cordova}}}]{garani2023JCAP...07..038G}
\bibinfo{author}{\bibfnamefont{R.}~\bibnamefont{{Garani}}},
  \bibinfo{author}{\bibfnamefont{N.}~\bibnamefont{{Raj}}}, \bibnamefont{and}
  \bibinfo{author}{\bibfnamefont{J.}~\bibnamefont{{Reynoso-Cordova}}},
  \bibinfo{journal}{\jcap} \textbf{\bibinfo{volume}{2023}}, \bibinfo{eid}{038}
  (\bibinfo{year}{2023}), \eprint{2303.18009}.

\bibitem[{\citenamefont{{Weinberg} et~al.}(2015)\citenamefont{{Weinberg},
  {Bullock}, {Governato}, {Kuzio de Naray}, and {Peter}}}]{2015PNAS..11212249W}
\bibinfo{author}{\bibfnamefont{D.~H.} \bibnamefont{{Weinberg}}},
  \bibinfo{author}{\bibfnamefont{J.~S.} \bibnamefont{{Bullock}}},
  \bibinfo{author}{\bibfnamefont{F.}~\bibnamefont{{Governato}}},
  \bibinfo{author}{\bibfnamefont{R.}~\bibnamefont{{Kuzio de Naray}}},
  \bibnamefont{and} \bibinfo{author}{\bibfnamefont{A.~H.~G.}
  \bibnamefont{{Peter}}}, \bibinfo{journal}{Proceedings of the National Academy
  of Science} \textbf{\bibinfo{volume}{112}}, \bibinfo{pages}{12249}
  (\bibinfo{year}{2015}), \eprint{1306.0913}.

\bibitem[{\citenamefont{Mosbech et~al.}(2023)\citenamefont{Mosbech, Jenkins,
  Bose, Boehm, Sakellariadou, and Wong}}]{PhysRevD.108.043512}
\bibinfo{author}{\bibfnamefont{M.~R.} \bibnamefont{Mosbech}},
  \bibinfo{author}{\bibfnamefont{A.~C.} \bibnamefont{Jenkins}},
  \bibinfo{author}{\bibfnamefont{S.}~\bibnamefont{Bose}},
  \bibinfo{author}{\bibfnamefont{C.}~\bibnamefont{Boehm}},
  \bibinfo{author}{\bibfnamefont{M.}~\bibnamefont{Sakellariadou}},
  \bibnamefont{and} \bibinfo{author}{\bibfnamefont{Y.~Y.~Y.}
  \bibnamefont{Wong}}, \bibinfo{journal}{Phys. Rev. D}
  \textbf{\bibinfo{volume}{108}}, \bibinfo{pages}{043512}
  (\bibinfo{year}{2023}),
  \urlprefix\url{https://link.aps.org/doi/10.1103/PhysRevD.108.043512}.

\bibitem[{\citenamefont{Fornal et~al.}(2023{\natexlab{a}})\citenamefont{Fornal,
  Garcia, and Pierre}}]{fornalPhysRevD.108.055022}
\bibinfo{author}{\bibfnamefont{B.}~\bibnamefont{Fornal}},
  \bibinfo{author}{\bibfnamefont{K.}~\bibnamefont{Garcia}}, \bibnamefont{and}
  \bibinfo{author}{\bibfnamefont{E.}~\bibnamefont{Pierre}},
  \bibinfo{journal}{Phys. Rev. D} \textbf{\bibinfo{volume}{108}},
  \bibinfo{pages}{055022} (\bibinfo{year}{2023}{\natexlab{a}}),
  \urlprefix\url{https://link.aps.org/doi/10.1103/PhysRevD.108.055022}.

\bibitem[{\citenamefont{Samanta and Urban}(2022)}]{Samanta_2022}
\bibinfo{author}{\bibfnamefont{R.}~\bibnamefont{Samanta}} \bibnamefont{and}
  \bibinfo{author}{\bibfnamefont{F.~R.} \bibnamefont{Urban}},
  \bibinfo{journal}{Journal of Cosmology and Astroparticle Physics}
  \textbf{\bibinfo{volume}{2022}}, \bibinfo{pages}{017} (\bibinfo{year}{2022}),
  \urlprefix\url{https://dx.doi.org/10.1088/1475-7516/2022/06/017}.

\bibitem[{\citenamefont{Fornal et~al.}(2023{\natexlab{b}})\citenamefont{Fornal,
  Garcia, and Pierre}}]{mosbePhysRevD.108.055022}
\bibinfo{author}{\bibfnamefont{B.}~\bibnamefont{Fornal}},
  \bibinfo{author}{\bibfnamefont{K.}~\bibnamefont{Garcia}}, \bibnamefont{and}
  \bibinfo{author}{\bibfnamefont{E.}~\bibnamefont{Pierre}},
  \bibinfo{journal}{Phys. Rev. D} \textbf{\bibinfo{volume}{108}},
  \bibinfo{pages}{055022} (\bibinfo{year}{2023}{\natexlab{b}}),
  \urlprefix\url{https://link.aps.org/doi/10.1103/PhysRevD.108.055022}.

\bibitem[{\citenamefont{Banerjee et~al.}(2023)\citenamefont{Banerjee, Bera, and
  Mota}}]{Banerjee_2023}
\bibinfo{author}{\bibfnamefont{S.}~\bibnamefont{Banerjee}},
  \bibinfo{author}{\bibfnamefont{S.}~\bibnamefont{Bera}}, \bibnamefont{and}
  \bibinfo{author}{\bibfnamefont{D.~F.} \bibnamefont{Mota}},
  \bibinfo{journal}{Journal of Cosmology and Astroparticle Physics}
  \textbf{\bibinfo{volume}{2023}}, \bibinfo{pages}{041} (\bibinfo{year}{2023}),
  \urlprefix\url{https://dx.doi.org/10.1088/1475-7516/2023/03/041}.

\bibitem[{\citenamefont{Ghoshal and Strumia}(2023)}]{ghoshal2023probing}
\bibinfo{author}{\bibfnamefont{A.}~\bibnamefont{Ghoshal}} \bibnamefont{and}
  \bibinfo{author}{\bibfnamefont{A.}~\bibnamefont{Strumia}},
  \emph{\bibinfo{title}{Probing the dark matter density with gravitational
  waves from super-massive binary black holes}} (\bibinfo{year}{2023}),
  \eprint{2306.17158}.

\bibitem[{\citenamefont{Mishra}(2022)}]{Mishra2022-ii}
\bibinfo{author}{\bibfnamefont{A.~K.} \bibnamefont{Mishra}},
  \bibinfo{journal}{The European Physical Journal C}
  \textbf{\bibinfo{volume}{82}}, \bibinfo{pages}{1060} (\bibinfo{year}{2022}).

\bibitem[{\citenamefont{Lu et~al.}(2018)\citenamefont{Lu, Huang, Wu, and
  Zhou}}]{lu2018damping}
\bibinfo{author}{\bibfnamefont{B.-Q.} \bibnamefont{Lu}},
  \bibinfo{author}{\bibfnamefont{D.}~\bibnamefont{Huang}},
  \bibinfo{author}{\bibfnamefont{Y.-L.} \bibnamefont{Wu}}, \bibnamefont{and}
  \bibinfo{author}{\bibfnamefont{Y.-F.} \bibnamefont{Zhou}},
  \emph{\bibinfo{title}{Damping of gravitational waves in a viscous universe
  and its implication for dark matter self-interactions}}
  (\bibinfo{year}{2018}), \eprint{1803.11397}.

\bibitem[{\citenamefont{Brevik and
  Nojiri}(2019)}]{gwviscodoi:10.1142/S0218271819501335}
\bibinfo{author}{\bibfnamefont{I.}~\bibnamefont{Brevik}} \bibnamefont{and}
  \bibinfo{author}{\bibfnamefont{S.}~\bibnamefont{Nojiri}},
  \bibinfo{journal}{International Journal of Modern Physics D}
  \textbf{\bibinfo{volume}{28}}, \bibinfo{pages}{1950133}
  (\bibinfo{year}{2019}), \eprint{https://doi.org/10.1142/S0218271819501335},
  \urlprefix\url{https://doi.org/10.1142/S0218271819501335}.

\bibitem[{\citenamefont{Dietrich et~al.}(2018)\citenamefont{Dietrich, Radice,
  Bernuzzi, Zappa, Perego, Brügmann, Chaurasia, Dudi, Tichy, and
  Ujevic}}]{Dietrich_2018}
\bibinfo{author}{\bibfnamefont{T.}~\bibnamefont{Dietrich}},
  \bibinfo{author}{\bibfnamefont{D.}~\bibnamefont{Radice}},
  \bibinfo{author}{\bibfnamefont{S.}~\bibnamefont{Bernuzzi}},
  \bibinfo{author}{\bibfnamefont{F.}~\bibnamefont{Zappa}},
  \bibinfo{author}{\bibfnamefont{A.}~\bibnamefont{Perego}},
  \bibinfo{author}{\bibfnamefont{B.}~\bibnamefont{Brügmann}},
  \bibinfo{author}{\bibfnamefont{S.~V.} \bibnamefont{Chaurasia}},
  \bibinfo{author}{\bibfnamefont{R.}~\bibnamefont{Dudi}},
  \bibinfo{author}{\bibfnamefont{W.}~\bibnamefont{Tichy}}, \bibnamefont{and}
  \bibinfo{author}{\bibfnamefont{M.}~\bibnamefont{Ujevic}},
  \bibinfo{journal}{Classical and Quantum Gravity}
  \textbf{\bibinfo{volume}{35}}, \bibinfo{pages}{24LT01}
  (\bibinfo{year}{2018}),
  \urlprefix\url{https://dx.doi.org/10.1088/1361-6382/aaebc0}.

\bibitem[{\citenamefont{Radice et~al.}(2020)\citenamefont{Radice, Bernuzzi, and
  Perego}}]{annurev:/content/journals/10.1146/annurev-nucl-013120-114541}
\bibinfo{author}{\bibfnamefont{D.}~\bibnamefont{Radice}},
  \bibinfo{author}{\bibfnamefont{S.}~\bibnamefont{Bernuzzi}}, \bibnamefont{and}
  \bibinfo{author}{\bibfnamefont{A.}~\bibnamefont{Perego}},
  \bibinfo{journal}{Annual Review of Nuclear and Particle Science}
  \textbf{\bibinfo{volume}{70}}, \bibinfo{pages}{95} (\bibinfo{year}{2020}),
  ISSN \bibinfo{issn}{1545-4134},
  \urlprefix\url{https://www.annualreviews.org/content/journals/10.1146/annurev-nucl-013120-114541}.

\bibitem[{\citenamefont{Foucart et~al.}(2024)\citenamefont{Foucart, Duez,
  Kidder, Pfeiffer, and Scheel}}]{bnssimPhysRevD.110.024003}
\bibinfo{author}{\bibfnamefont{F.}~\bibnamefont{Foucart}},
  \bibinfo{author}{\bibfnamefont{M.~D.} \bibnamefont{Duez}},
  \bibinfo{author}{\bibfnamefont{L.~E.} \bibnamefont{Kidder}},
  \bibinfo{author}{\bibfnamefont{H.~P.} \bibnamefont{Pfeiffer}},
  \bibnamefont{and} \bibinfo{author}{\bibfnamefont{M.~A.}
  \bibnamefont{Scheel}}, \bibinfo{journal}{Phys. Rev. D}
  \textbf{\bibinfo{volume}{110}}, \bibinfo{pages}{024003}
  (\bibinfo{year}{2024}),
  \urlprefix\url{https://link.aps.org/doi/10.1103/PhysRevD.110.024003}.

\bibitem[{\citenamefont{Nedora et~al.}(2021)\citenamefont{Nedora, Bernuzzi,
  Radice, Daszuta, Endrizzi, Perego, Prakash, Safarzadeh, Schianchi, and
  Logoteta}}]{Nedora_2021}
\bibinfo{author}{\bibfnamefont{V.}~\bibnamefont{Nedora}},
  \bibinfo{author}{\bibfnamefont{S.}~\bibnamefont{Bernuzzi}},
  \bibinfo{author}{\bibfnamefont{D.}~\bibnamefont{Radice}},
  \bibinfo{author}{\bibfnamefont{B.}~\bibnamefont{Daszuta}},
  \bibinfo{author}{\bibfnamefont{A.}~\bibnamefont{Endrizzi}},
  \bibinfo{author}{\bibfnamefont{A.}~\bibnamefont{Perego}},
  \bibinfo{author}{\bibfnamefont{A.}~\bibnamefont{Prakash}},
  \bibinfo{author}{\bibfnamefont{M.}~\bibnamefont{Safarzadeh}},
  \bibinfo{author}{\bibfnamefont{F.}~\bibnamefont{Schianchi}},
  \bibnamefont{and} \bibinfo{author}{\bibfnamefont{D.}~\bibnamefont{Logoteta}},
  \bibinfo{journal}{The Astrophysical Journal} \textbf{\bibinfo{volume}{906}},
  \bibinfo{pages}{98} (\bibinfo{year}{2021}),
  \urlprefix\url{https://dx.doi.org/10.3847/1538-4357/abc9be}.

\bibitem[{\citenamefont{{De Pietri} et~al.}(2020)\citenamefont{{De Pietri},
  {Feo}, {Font}, {L{\"o}ffler}, {Pasquali}, and
  {Stergioulas}}}]{Depietri2020PhRvD.101f4052D}
\bibinfo{author}{\bibfnamefont{R.}~\bibnamefont{{De Pietri}}},
  \bibinfo{author}{\bibfnamefont{A.}~\bibnamefont{{Feo}}},
  \bibinfo{author}{\bibfnamefont{J.~A.} \bibnamefont{{Font}}},
  \bibinfo{author}{\bibfnamefont{F.}~\bibnamefont{{L{\"o}ffler}}},
  \bibinfo{author}{\bibfnamefont{M.}~\bibnamefont{{Pasquali}}},
  \bibnamefont{and}
  \bibinfo{author}{\bibfnamefont{N.}~\bibnamefont{{Stergioulas}}},
  \bibinfo{journal}{\prd} \textbf{\bibinfo{volume}{101}}, \bibinfo{eid}{064052}
  (\bibinfo{year}{2020}), \eprint{1910.04036}.

\bibitem[{\citenamefont{Topolski et~al.}(2023)\citenamefont{Topolski, Tootle,
  and Rezzolla}}]{Topolski_2024}
\bibinfo{author}{\bibfnamefont{K.}~\bibnamefont{Topolski}},
  \bibinfo{author}{\bibfnamefont{S.~D.} \bibnamefont{Tootle}},
  \bibnamefont{and} \bibinfo{author}{\bibfnamefont{L.}~\bibnamefont{Rezzolla}},
  \bibinfo{journal}{The Astrophysical Journal} \textbf{\bibinfo{volume}{960}},
  \bibinfo{pages}{86} (\bibinfo{year}{2023}),
  \urlprefix\url{https://dx.doi.org/10.3847/1538-4357/ad0152}.

\bibitem[{\citenamefont{Bell et~al.}(2024)\citenamefont{Bell, Busoni, Robles,
  and Virgato}}]{Bell_2024}
\bibinfo{author}{\bibfnamefont{N.~F.} \bibnamefont{Bell}},
  \bibinfo{author}{\bibfnamefont{G.}~\bibnamefont{Busoni}},
  \bibinfo{author}{\bibfnamefont{S.}~\bibnamefont{Robles}}, \bibnamefont{and}
  \bibinfo{author}{\bibfnamefont{M.}~\bibnamefont{Virgato}},
  \bibinfo{journal}{Journal of Cosmology and Astroparticle Physics}
  \textbf{\bibinfo{volume}{2024}}, \bibinfo{pages}{006} (\bibinfo{year}{2024}),
  ISSN \bibinfo{issn}{1475-7516},
  \urlprefix\url{http://dx.doi.org/10.1088/1475-7516/2024/04/006}.

\bibitem[{\citenamefont{Herrero et~al.}(2019)\citenamefont{Herrero,
  P\'erez-Garc\'{\i}a, Silk, and Albertus}}]{herreroPhysRevD.100.103019}
\bibinfo{author}{\bibfnamefont{A.}~\bibnamefont{Herrero}},
  \bibinfo{author}{\bibfnamefont{M.~A.} \bibnamefont{P\'erez-Garc\'{\i}a}},
  \bibinfo{author}{\bibfnamefont{J.}~\bibnamefont{Silk}}, \bibnamefont{and}
  \bibinfo{author}{\bibfnamefont{C.}~\bibnamefont{Albertus}},
  \bibinfo{journal}{Phys. Rev. D} \textbf{\bibinfo{volume}{100}},
  \bibinfo{pages}{103019} (\bibinfo{year}{2019}),
  \urlprefix\url{https://link.aps.org/doi/10.1103/PhysRevD.100.103019}.

\bibitem[{\citenamefont{Kiuchi et~al.}(2020)\citenamefont{Kiuchi, Kawaguchi,
  Kyutoku, Sekiguchi, and Shibata}}]{KiuchiPhysRevD.101.084006}
\bibinfo{author}{\bibfnamefont{K.}~\bibnamefont{Kiuchi}},
  \bibinfo{author}{\bibfnamefont{K.}~\bibnamefont{Kawaguchi}},
  \bibinfo{author}{\bibfnamefont{K.}~\bibnamefont{Kyutoku}},
  \bibinfo{author}{\bibfnamefont{Y.}~\bibnamefont{Sekiguchi}},
  \bibnamefont{and} \bibinfo{author}{\bibfnamefont{M.}~\bibnamefont{Shibata}},
  \bibinfo{journal}{Phys. Rev. D} \textbf{\bibinfo{volume}{101}},
  \bibinfo{pages}{084006} (\bibinfo{year}{2020}),
  \urlprefix\url{https://link.aps.org/doi/10.1103/PhysRevD.101.084006}.

\bibitem[{\citenamefont{Bhatt et~al.}(2019)\citenamefont{Bhatt, Mishra, and
  Nayak}}]{Bhatt_2019}
\bibinfo{author}{\bibfnamefont{J.~R.} \bibnamefont{Bhatt}},
  \bibinfo{author}{\bibfnamefont{A.~K.} \bibnamefont{Mishra}},
  \bibnamefont{and} \bibinfo{author}{\bibfnamefont{A.~C.} \bibnamefont{Nayak}},
  \bibinfo{journal}{Physical Review D} \textbf{\bibinfo{volume}{100}}
  (\bibinfo{year}{2019}), ISSN \bibinfo{issn}{2470-0029},
  \urlprefix\url{http://dx.doi.org/10.1103/PhysRevD.100.063539}.

\bibitem[{\citenamefont{Alonso-\'Alvarez
  et~al.}(2024)\citenamefont{Alonso-\'Alvarez, Cline, and
  Dewar}}]{GonzaloPhysRevLett.133.021401}
\bibinfo{author}{\bibfnamefont{G.}~\bibnamefont{Alonso-\'Alvarez}},
  \bibinfo{author}{\bibfnamefont{J.~M.} \bibnamefont{Cline}}, \bibnamefont{and}
  \bibinfo{author}{\bibfnamefont{C.}~\bibnamefont{Dewar}},
  \bibinfo{journal}{Phys. Rev. Lett.} \textbf{\bibinfo{volume}{133}},
  \bibinfo{pages}{021401} (\bibinfo{year}{2024}),
  \urlprefix\url{https://link.aps.org/doi/10.1103/PhysRevLett.133.021401}.

\bibitem[{\citenamefont{Zappa et~al.}(2023)\citenamefont{Zappa, Bernuzzi,
  Radice, and Perego}}]{Zappa_2023}
\bibinfo{author}{\bibfnamefont{F.}~\bibnamefont{Zappa}},
  \bibinfo{author}{\bibfnamefont{S.}~\bibnamefont{Bernuzzi}},
  \bibinfo{author}{\bibfnamefont{D.}~\bibnamefont{Radice}}, \bibnamefont{and}
  \bibinfo{author}{\bibfnamefont{A.}~\bibnamefont{Perego}},
  \bibinfo{journal}{Monthly Notices of the Royal Astronomical Society}
  \textbf{\bibinfo{volume}{520}}, \bibinfo{pages}{1481–1503}
  (\bibinfo{year}{2023}), ISSN \bibinfo{issn}{1365-2966},
  \urlprefix\url{http://dx.doi.org/10.1093/mnras/stad107}.

\bibitem[{\citenamefont{Espino et~al.}(2023)\citenamefont{Espino, Hammond,
  Radice, Bernuzzi, Gamba, Zappa, Micchi, and
  Perego}}]{espino2023neutrinotrappingoutofequilibriumeffects}
\bibinfo{author}{\bibfnamefont{P.~L.} \bibnamefont{Espino}},
  \bibinfo{author}{\bibfnamefont{P.}~\bibnamefont{Hammond}},
  \bibinfo{author}{\bibfnamefont{D.}~\bibnamefont{Radice}},
  \bibinfo{author}{\bibfnamefont{S.}~\bibnamefont{Bernuzzi}},
  \bibinfo{author}{\bibfnamefont{R.}~\bibnamefont{Gamba}},
  \bibinfo{author}{\bibfnamefont{F.}~\bibnamefont{Zappa}},
  \bibinfo{author}{\bibfnamefont{L.~F.~L.} \bibnamefont{Micchi}},
  \bibnamefont{and} \bibinfo{author}{\bibfnamefont{A.}~\bibnamefont{Perego}},
  \emph{\bibinfo{title}{Neutrino trapping and out-of-equilibrium effects in
  binary neutron star merger remnants}} (\bibinfo{year}{2023}),
  \eprint{2311.00031}, \urlprefix\url{https://arxiv.org/abs/2311.00031}.

\bibitem[{\citenamefont{{Banik} et~al.}(2014)\citenamefont{{Banik}, {Hempel},
  and {Bandyopadhyay}}}]{2014ApJS..214...22B}
\bibinfo{author}{\bibfnamefont{S.}~\bibnamefont{{Banik}}},
  \bibinfo{author}{\bibfnamefont{M.}~\bibnamefont{{Hempel}}}, \bibnamefont{and}
  \bibinfo{author}{\bibfnamefont{D.}~\bibnamefont{{Bandyopadhyay}}},
  \bibinfo{journal}{\apjs} \textbf{\bibinfo{volume}{214}}, \bibinfo{eid}{22}
  (\bibinfo{year}{2014}), \eprint{1404.6173}.

\bibitem[{\citenamefont{Steiner et~al.}(2013)\citenamefont{Steiner, Hempel, and
  Fischer}}]{steiner2013core}
\bibinfo{author}{\bibfnamefont{A.~W.} \bibnamefont{Steiner}},
  \bibinfo{author}{\bibfnamefont{M.}~\bibnamefont{Hempel}}, \bibnamefont{and}
  \bibinfo{author}{\bibfnamefont{T.}~\bibnamefont{Fischer}},
  \bibinfo{journal}{The Astrophysical Journal} \textbf{\bibinfo{volume}{774}},
  \bibinfo{pages}{17} (\bibinfo{year}{2013}).

\bibitem[{\citenamefont{{Typel} et~al.}(2010)\citenamefont{{Typel},
  {R{\"o}pke}, {Kl{\"a}hn}, {Blaschke}, and {Wolter}}}]{2010PhRvC..81a5803T}
\bibinfo{author}{\bibfnamefont{S.}~\bibnamefont{{Typel}}},
  \bibinfo{author}{\bibfnamefont{G.}~\bibnamefont{{R{\"o}pke}}},
  \bibinfo{author}{\bibfnamefont{T.}~\bibnamefont{{Kl{\"a}hn}}},
  \bibinfo{author}{\bibfnamefont{D.}~\bibnamefont{{Blaschke}}},
  \bibnamefont{and} \bibinfo{author}{\bibfnamefont{H.~H.}
  \bibnamefont{{Wolter}}}, \bibinfo{journal}{\prc}
  \textbf{\bibinfo{volume}{81}}, \bibinfo{eid}{015803} (\bibinfo{year}{2010}),
  \eprint{0908.2344}.

\bibitem[{\citenamefont{Prada et~al.}(2020)\citenamefont{Prada, Content,
  Goobar, Izzo, Pérez, Agnello, del Burgo, Dhillon, Diego, Galbany
  et~al.}}]{prada2020whitepapermaatgtc}
\bibinfo{author}{\bibfnamefont{F.}~\bibnamefont{Prada}},
  \bibinfo{author}{\bibfnamefont{R.}~\bibnamefont{Content}},
  \bibinfo{author}{\bibfnamefont{A.}~\bibnamefont{Goobar}},
  \bibinfo{author}{\bibfnamefont{L.}~\bibnamefont{Izzo}},
  \bibinfo{author}{\bibfnamefont{E.}~\bibnamefont{Pérez}},
  \bibinfo{author}{\bibfnamefont{A.}~\bibnamefont{Agnello}},
  \bibinfo{author}{\bibfnamefont{C.}~\bibnamefont{del Burgo}},
  \bibinfo{author}{\bibfnamefont{V.}~\bibnamefont{Dhillon}},
  \bibinfo{author}{\bibfnamefont{J.~M.} \bibnamefont{Diego}},
  \bibinfo{author}{\bibfnamefont{L.}~\bibnamefont{Galbany}},
  \bibnamefont{et~al.}, \emph{\bibinfo{title}{White paper on maat@gtc}}
  (\bibinfo{year}{2020}), \eprint{2007.01603},
  \urlprefix\url{https://arxiv.org/abs/2007.01603}.

\bibitem[{\citenamefont{Bezares et~al.}(2019)\citenamefont{Bezares, Vigan\`o,
  and Palenzuela}}]{bezaresPhysRevD.100.044049}
\bibinfo{author}{\bibfnamefont{M.}~\bibnamefont{Bezares}},
  \bibinfo{author}{\bibfnamefont{D.}~\bibnamefont{Vigan\`o}}, \bibnamefont{and}
  \bibinfo{author}{\bibfnamefont{C.}~\bibnamefont{Palenzuela}},
  \bibinfo{journal}{Phys. Rev. D} \textbf{\bibinfo{volume}{100}},
  \bibinfo{pages}{044049} (\bibinfo{year}{2019}),
  \urlprefix\url{https://link.aps.org/doi/10.1103/PhysRevD.100.044049}.

\bibitem[{\citenamefont{Kastaun and
  Galeazzi}(2015)}]{KastaunPhysRevD.91.064027}
\bibinfo{author}{\bibfnamefont{W.}~\bibnamefont{Kastaun}} \bibnamefont{and}
  \bibinfo{author}{\bibfnamefont{F.}~\bibnamefont{Galeazzi}},
  \bibinfo{journal}{Phys. Rev. D} \textbf{\bibinfo{volume}{91}},
  \bibinfo{pages}{064027} (\bibinfo{year}{2015}),
  \urlprefix\url{https://link.aps.org/doi/10.1103/PhysRevD.91.064027}.

\bibitem[{\citenamefont{Glampedakis and Gualtieri}(2018)}]{Glampedakis_2018}
\bibinfo{author}{\bibfnamefont{K.}~\bibnamefont{Glampedakis}} \bibnamefont{and}
  \bibinfo{author}{\bibfnamefont{L.}~\bibnamefont{Gualtieri}},
  \emph{\bibinfo{title}{Gravitational Waves from Single Neutron Stars: An
  Advanced Detector Era Survey}} (\bibinfo{publisher}{Springer International
  Publishing}, \bibinfo{year}{2018}), p. \bibinfo{pages}{673–736}, ISBN
  \bibinfo{isbn}{9783319976167},
  \urlprefix\url{http://dx.doi.org/10.1007/978-3-319-97616-7_12}.

\bibitem[{\citenamefont{Dall’Osso et~al.}(2018)\citenamefont{Dall’Osso,
  Stella, and Palomba}}]{Dall_Osso_2018}
\bibinfo{author}{\bibfnamefont{S.}~\bibnamefont{Dall’Osso}},
  \bibinfo{author}{\bibfnamefont{L.}~\bibnamefont{Stella}}, \bibnamefont{and}
  \bibinfo{author}{\bibfnamefont{C.}~\bibnamefont{Palomba}},
  \bibinfo{journal}{Monthly Notices of the Royal Astronomical Society}
  \textbf{\bibinfo{volume}{480}}, \bibinfo{pages}{1353–1362}
  (\bibinfo{year}{2018}), ISSN \bibinfo{issn}{1365-2966},
  \urlprefix\url{http://dx.doi.org/10.1093/mnras/sty1706}.

\bibitem[{\citenamefont{Thorne et~al.}(2000)\citenamefont{Thorne, Misner, and
  Wheeler}}]{thorne2000gravitation}
\bibinfo{author}{\bibfnamefont{K.~S.} \bibnamefont{Thorne}},
  \bibinfo{author}{\bibfnamefont{C.~W.} \bibnamefont{Misner}},
  \bibnamefont{and} \bibinfo{author}{\bibfnamefont{J.~A.}
  \bibnamefont{Wheeler}}, \emph{\bibinfo{title}{Gravitation}}
  (\bibinfo{publisher}{Freeman San Francisco, CA}, \bibinfo{year}{2000}).

\bibitem[{\citenamefont{{Radice} and {Rezzolla}}(2012)}]{2012A&A...547A..26R}
\bibinfo{author}{\bibfnamefont{D.}~\bibnamefont{{Radice}}} \bibnamefont{and}
  \bibinfo{author}{\bibfnamefont{L.}~\bibnamefont{{Rezzolla}}},
  \bibinfo{journal}{\aap} \textbf{\bibinfo{volume}{547}}, \bibinfo{eid}{A26}
  (\bibinfo{year}{2012}), \eprint{1206.6502}.

\bibitem[{\citenamefont{Alford et~al.}(2018)\citenamefont{Alford, Bovard,
  Hanauske, Rezzolla, and Schwenzer}}]{Alford_2018}
\bibinfo{author}{\bibfnamefont{M.~G.} \bibnamefont{Alford}},
  \bibinfo{author}{\bibfnamefont{L.}~\bibnamefont{Bovard}},
  \bibinfo{author}{\bibfnamefont{M.}~\bibnamefont{Hanauske}},
  \bibinfo{author}{\bibfnamefont{L.}~\bibnamefont{Rezzolla}}, \bibnamefont{and}
  \bibinfo{author}{\bibfnamefont{K.}~\bibnamefont{Schwenzer}},
  \bibinfo{journal}{Physical Review Letters} \textbf{\bibinfo{volume}{120}}
  (\bibinfo{year}{2018}), ISSN \bibinfo{issn}{1079-7114},
  \urlprefix\url{http://dx.doi.org/10.1103/PhysRevLett.120.041101}.

\bibitem[{\citenamefont{Alford et~al.}(2022)\citenamefont{Alford, Harutyunyan,
  and Sedrakian}}]{Alford_2022}
\bibinfo{author}{\bibfnamefont{M.}~\bibnamefont{Alford}},
  \bibinfo{author}{\bibfnamefont{A.}~\bibnamefont{Harutyunyan}},
  \bibnamefont{and}
  \bibinfo{author}{\bibfnamefont{A.}~\bibnamefont{Sedrakian}},
  \bibinfo{journal}{Particles} \textbf{\bibinfo{volume}{5}},
  \bibinfo{pages}{361–376} (\bibinfo{year}{2022}), ISSN
  \bibinfo{issn}{2571-712X},
  \urlprefix\url{http://dx.doi.org/10.3390/particles5030029}.

\bibitem[{\citenamefont{Radice et~al.}(2022)\citenamefont{Radice, Bernuzzi,
  Perego, and Haas}}]{Radice_2022}
\bibinfo{author}{\bibfnamefont{D.}~\bibnamefont{Radice}},
  \bibinfo{author}{\bibfnamefont{S.}~\bibnamefont{Bernuzzi}},
  \bibinfo{author}{\bibfnamefont{A.}~\bibnamefont{Perego}}, \bibnamefont{and}
  \bibinfo{author}{\bibfnamefont{R.}~\bibnamefont{Haas}},
  \bibinfo{journal}{Monthly Notices of the Royal Astronomical Society}
  \textbf{\bibinfo{volume}{512}}, \bibinfo{pages}{1499–1521}
  (\bibinfo{year}{2022}), ISSN \bibinfo{issn}{1365-2966},
  \urlprefix\url{http://dx.doi.org/10.1093/mnras/stac589}.

\bibitem[{\citenamefont{Foreman-Mackey
  et~al.}(2013)\citenamefont{Foreman-Mackey, Hogg, Lang, and
  Goodman}}]{Foreman_Mackey_2013}
\bibinfo{author}{\bibfnamefont{D.}~\bibnamefont{Foreman-Mackey}},
  \bibinfo{author}{\bibfnamefont{D.~W.} \bibnamefont{Hogg}},
  \bibinfo{author}{\bibfnamefont{D.}~\bibnamefont{Lang}}, \bibnamefont{and}
  \bibinfo{author}{\bibfnamefont{J.}~\bibnamefont{Goodman}},
  \bibinfo{journal}{Publications of the Astronomical Society of the Pacific}
  \textbf{\bibinfo{volume}{125}}, \bibinfo{pages}{306–312}
  (\bibinfo{year}{2013}), ISSN \bibinfo{issn}{1538-3873},
  \urlprefix\url{http://dx.doi.org/10.1086/670067}.

\bibitem[{\citenamefont{Anand et~al.}(2018)\citenamefont{Anand, Chaubal,
  Mazumdar, Mohanty, and Parashari}}]{Anand_2018}
\bibinfo{author}{\bibfnamefont{S.}~\bibnamefont{Anand}},
  \bibinfo{author}{\bibfnamefont{P.}~\bibnamefont{Chaubal}},
  \bibinfo{author}{\bibfnamefont{A.}~\bibnamefont{Mazumdar}},
  \bibinfo{author}{\bibfnamefont{S.}~\bibnamefont{Mohanty}}, \bibnamefont{and}
  \bibinfo{author}{\bibfnamefont{P.}~\bibnamefont{Parashari}},
  \bibinfo{journal}{Journal of Cosmology and Astroparticle Physics}
  \textbf{\bibinfo{volume}{2018}}, \bibinfo{pages}{031–031}
  (\bibinfo{year}{2018}), ISSN \bibinfo{issn}{1475-7516},
  \urlprefix\url{http://dx.doi.org/10.1088/1475-7516/2018/05/031}.

\bibitem[{\citenamefont{Soultanis et~al.}(2022)\citenamefont{Soultanis,
  Bauswein, and Stergioulas}}]{soultanis2022analytic}
\bibinfo{author}{\bibfnamefont{T.}~\bibnamefont{Soultanis}},
  \bibinfo{author}{\bibfnamefont{A.}~\bibnamefont{Bauswein}}, \bibnamefont{and}
  \bibinfo{author}{\bibfnamefont{N.}~\bibnamefont{Stergioulas}},
  \bibinfo{journal}{Physical Review D} \textbf{\bibinfo{volume}{105}},
  \bibinfo{pages}{043020} (\bibinfo{year}{2022}).

\bibitem[{\citenamefont{Fay}(2012)}]{fay2012coulomb}
\bibinfo{author}{\bibfnamefont{T.~H.} \bibnamefont{Fay}},
  \bibinfo{journal}{International Journal of Mathematical Education in Science
  and Technology} \textbf{\bibinfo{volume}{43}}, \bibinfo{pages}{923}
  (\bibinfo{year}{2012}).

\end{thebibliography}

\end{document}